\documentclass[preprint,12pt]{revtex4-1}

\usepackage{amssymb}
\usepackage{epsfig}
\usepackage{subcaption}
\usepackage{multirow}
\usepackage{floatrow}

\begin{document}

\title{Magnetic interactions in doped silicene for spintronics}

\author{Lucca Moraes Gomes}
\author{Andreia Luisa da Rosa}
\affiliation{Institute of Physics, Quantum Computational Materials Science Group, Campus Samambaia, Goiania, 74690-900, Goiás, Brazil}

\begin{abstract}
Silicon is a material whose technological
application is well established, and obtaining this material in nanostructured form increases its possibility of integration in current technology.  Silicene is a natural compatibility with current silicon-based electronics industry. Furthermore, doping is a technique that can be often  used to adjust the band gap of silicene and at the same time introduce new functions. Here we investigate Several aspects of silicene doping with chromium, such as dopant solubility limits, site preference for adsorption and doping, and formation of magnetic domains. In this work we carried out investigation on diffusion and doping of chromium atoms on silicene. We use density-functional theory to identify the electronic and magnetic properties of tchromium atoms on monolayer and bilayer silicene. We find that magnetization depends on key parameters such as buckling, interlayer distance and adsorption site.
\end{abstract}

\maketitle



\section{Introduction}
\label{sec:sample1}
The development of new experimental techniques and the emergence of powerful computers combined with iterative methods, gave rise to a new area of reesearch, allowing to explore low-dimensional materials with interesting properties that only can be seen and used by quantum mechanics. Those materials can have properties that could take humanity to next generation of technological devices because of their unique properties like high electronic mobility, thermal conductivity and impressive tenacity. Those properties may have application in several areas that goes from medical applications to the enhancement of new aircrafts (see for example \cite{artigo_01}).

Graphene is a low dimensional material that opened the new branch of condensed-matter-physics. The layered materials is made of carbon which a very reactive element of Group IV that have four half occupied orbitals. Graphene is a single layer of graphite, it has a honeycomb structure where each carbon atom bonds to other three atoms by $sp^2$ hybridization, hence leaving a half occupied orbital that is usually a $p_z$ orbital.  The band structure of graphene shows the so called Dirac cones. In this unique dispersion pattern, the electrons then behave as  Dirac fermions which are massless fermions.  This behavior in the band structure renders  a high mobility which could have an infinite number of electronic applications \cite{artigo_02}.

Two-dimensional carbon  was first proposed theoretically and had the first studied published in the beginning of the 60's by  Hans-Peter Boehm who studied extremely thin flakes of graphite and gave the name graphene for this flakes. The suffix "ene" is given because of the numerous double bond that thin layers have \cite{artigo_03}. Graphene was characterized only in 2004 by Geim and Novoselov, that granted them in 2010 the Nobel prize. They were able to obtain graphene through a  process called Scotch tape method which consist in simply applying the tape on thin flakes of graphite until one thin layer remain, using atomic force microscope they were able to identify the single layered material\cite{artigo_04}. 
Despite of the interesting properties of graphene, the manipulation of its electronic properties is very challenging \cite{artigo_05} which led to the the search for similar or better properties. 

At the same time, silicon is the most abundant element on earth crust, which is extremely explored in the  semiconductor industry.  Silicene, a two-dimensional silicon layer, became very targeted by researchers in the last few decades. The silicene structure resembles  the one of graphene\,\cite{artigo_07}. Silicene layers were synthesized on several substrates, such as Ag(111), Ir(111), ZrB2(0001), ZrC(111) and MoS$_2$ surfaces by epitaxial growth on those surfaces \cite{artigo_29,ARAFUNE2013}.
This has prompted experimental studies of the growth of multi-layer silicene, though the nature of its monolayered silicene structure has been investigated in a few papers\,\cite{Liu2013,Pawlak2019}. Like graphene, synthesizing free-standing silicene represents the ultimate challenge. A first step towards this has been reported recently through chemical exfoliation from calcium disilicide (CaSi2). 

 On the other hand, the discovery of atomic monolayer magnetic materials has triggered significant interest in the magnetism/spintronics and 2D van der Waals materials communities. The emergence of 2D magnetic materials presents a unique opportunity to study magnetism and spintronics devices in new regimes. This Review surveys the basic properties of these materials, methods to read and write their magnetic states, and emerging device concepts. There is a large class of layered magnetic materials with unique magnetic properties, which provides an ideal platform to study magnetism and spintronics device concepts in the 2D limit. Because these materials are atomically thin, their magnetic states can be effectively controlled or switched by external perturbations other than magnetic fields, such as electric fields, doping and strain. New materials concepts, such as magnetizing 2D semiconductors by magnetic proximity coupling and spin tunnel field-effect transistors are promising devices.

 Here we demonstrate that silicene decorated with certain 3d transition metals (Vanadium) can sustain a stable quantum anomalous Hall effect using both analytical model and first-principles Wannier interpolation\,\cite{Liu2013}. \,\cite{Sicluster2023}

In this work we investigating silicene to  Defects could be associated with vacancies in the nanosheet, adsorption of impurities in the sample or even the substitution of atoms. This work has the objective to study the changes on the properties by adsorbing open shell elements on silicene. The open shell element focused in this work is Chromium (Cr). Open shell elements adsorbed on silicene may have several application due to the modification of electronic and magnetic properties, which could render  application in the field of spintronics, where the spin is responsible for carrying the information\,\cite{artigo_09,artigo_10,artigo_12}.   
bits etc

\section{Methodology} 

We performed first-principles calculations  within the GGA (generalized-gradient approximation  to a better description of the electronic properties according to the 
parameterization of PBE (Perdew-Burke-Ernzerhof)\cite{artigo_20} for the exchange an correlation potential. The electrons wave-functions were built using the projected augmented wave PAW method \cite{artigo_23}. Silicene has an hexagonal symmetry with two atoms in the unit cell. The k-mesh used is (12x12x1), centered in $\Gamma$ point, according to Monkhorst-Pack\cite{artigo_26}. The single-electron wave-functions were expanded in plane-waves up to the energy cutoff of $400eV$. The technique used to calculate the diffusion barrier of Cr atoms on 
 silicene is known as NEB (Nudged Elastic Band) \cite{artigo_31}. WE have chosen 7 images to investigate single atom difusion on silicene.

\section{Results}

\subsection{Pristine silicene}


\begin{figure}[!htb]
\begin{center}
\includegraphics[scale=0.45,width=8cm,clip]{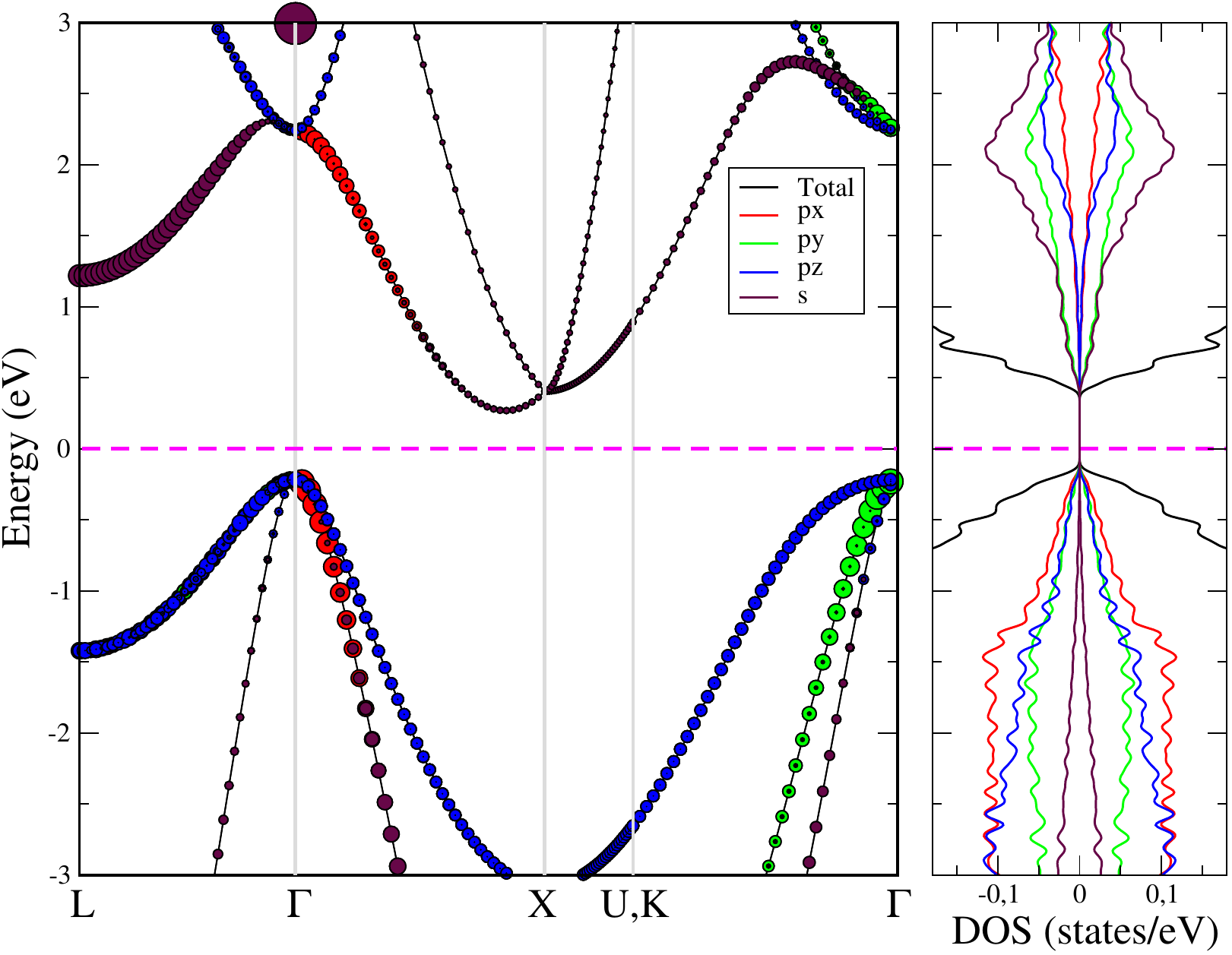}
\caption{Projected band structure and projected density of states for bulk silicon.}
\label{fig:band_bulk}
\end{center}
\end{figure}

\begin{figure}[!htb]
\centering
\includegraphics[scale=0.1,width = 6cm,clip]{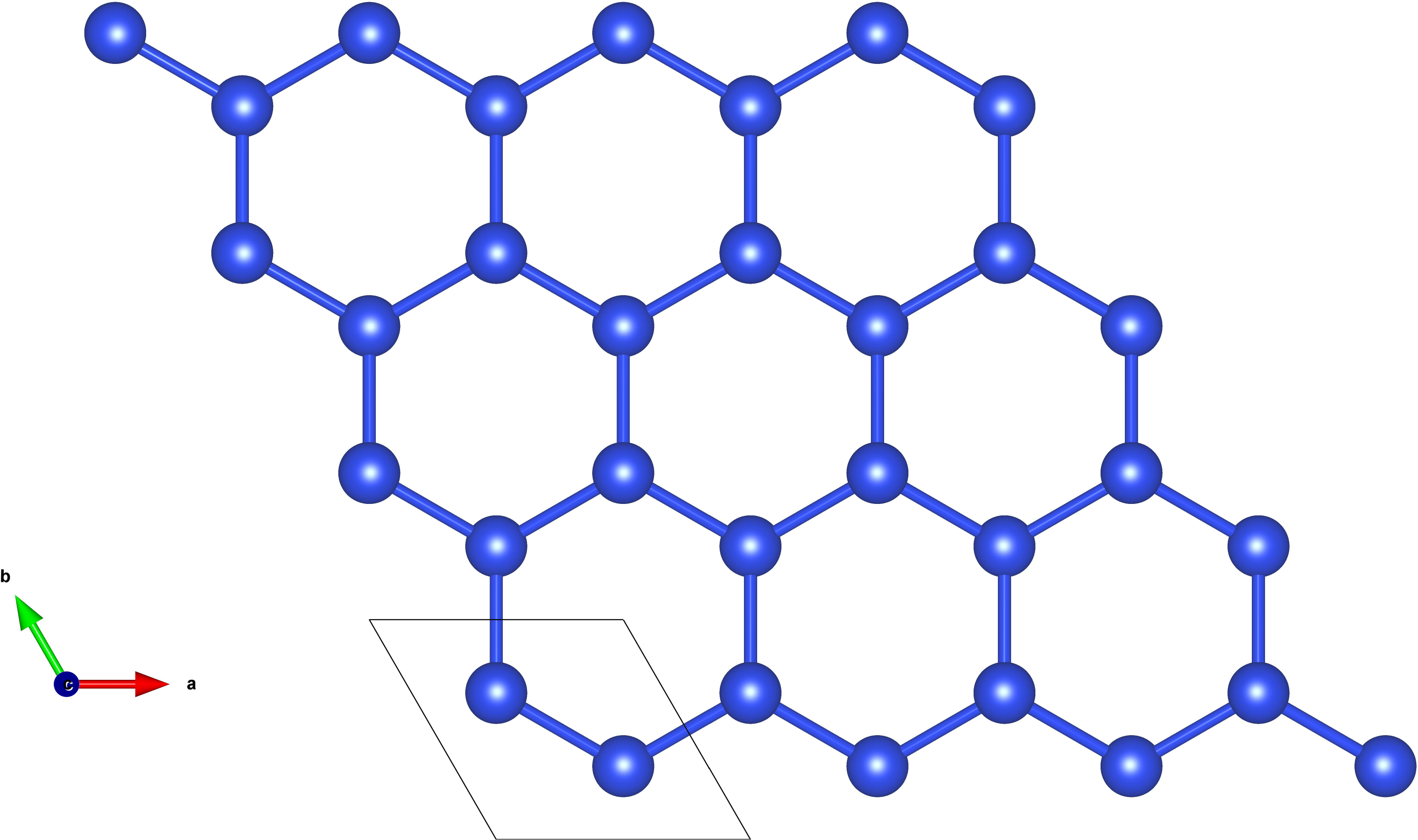}
\vspace{1cm}
\includegraphics[scale=0.02, width = 6cm,clip]{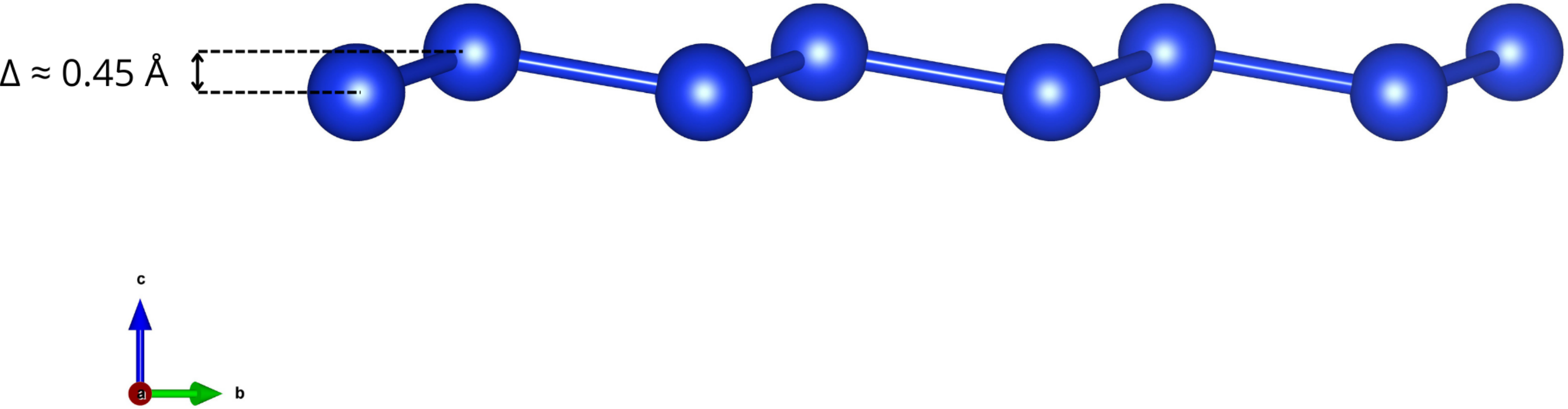}
\caption{Side and top views of buckled monolayer silicene}
\label{fig:buckled_silicene}
\end{figure}

One can obtain silicene  from bulk silicon by cutting  the bulk silicon along the (111) plane\,\cite{artigo_27}. Pristine silicene  was investigated in three different configuration: buckled, planar silicene and dumbbell silicene. The buckled silicene represents the ground state configuration for the two-dimensional silicon sheet. Buckled silicene as seen in Fig.\,\ref{fig:buckled_silicene}.The buckled state stability can be understood by orbital hybridization.

\begin{figure}[!htb]
    \centering
    \includegraphics[scale=0.5,width=7cm,clip]{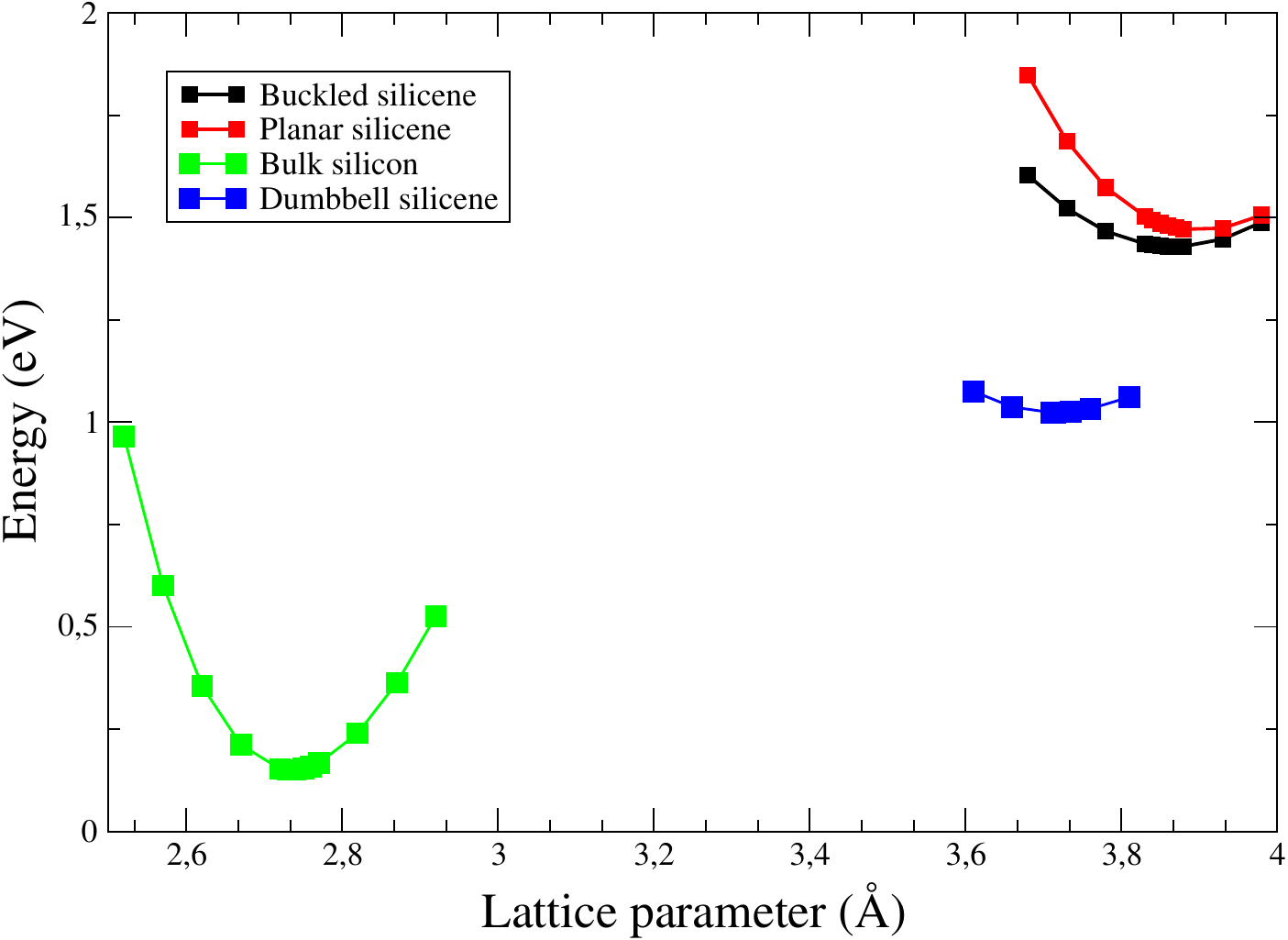}
    \caption{Relative total energy of bulk silicon and silicene as a function of the lattice parameter.}
    \label{fig:stability_silicene}
\end{figure}

The lattice parameter is $a$ = 3.87\AA, the buckling distance is $\Delta$ 0.45\AA, and the bond length between silicon atoms is $d$ = 2.28\AA. 

It is possible to see the indirect energy gap at the Fermi level of bulk silicon, the calculated energy gap  is E$_{gap}$ =0.46 eV. The experimental result for the energy gap is is 1.17 eV\,\cite{JAFARI2022}. Although the GGA underestimate the electronic band gap as shown in Figure\,\ref{fig:band_bulk}, the use of  GGA will be later justified. All the orbitals contribute to states in the top of the valence band and in the bottom of the conduction band. The spin up and spin down projected density of states is semiconductor.
  
Group IV atoms on a monolayer configuration tend to bond to other three atoms by $sp^2$ hybridization, forming $\sigma$ bonds, leaving a half occupied $p_z$ orbital perpendicular to the plane. What happens to the remaining orbital depends on which atom the structure is made of. A monolayer structure made of carbon will have adjacent unoccupied $p_z$ orbitals bounding to each other forming a strong $\pi$ bond due to the relative short distance between C-C interaction, resulting in a planar structure. Otherwise, if the structure is made of Silicon atoms the $\pi$ bond by adjacent $p_z$ orbital coupling is weakened by the relative long distance in Si-Si interaction. Thus, the $p_z$ orbital will prefer to combine with $s$ orbital forming a sp$^3$ state, resulting in structural modification \cite{artigo_07}.

\begin{figure}[!htb]
\centering
     \includegraphics[scale=0.45,width = 8cm,clip]{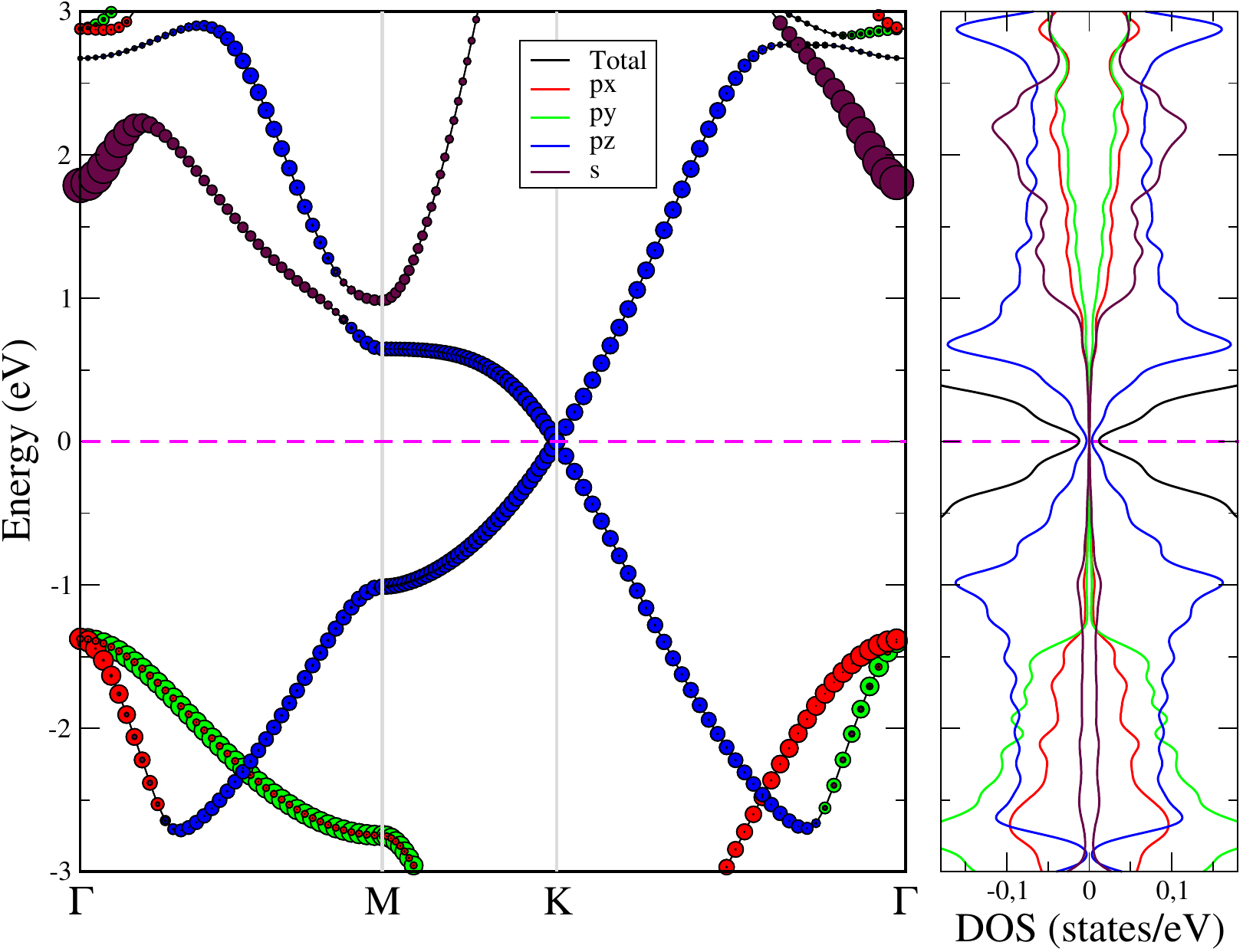}
    \caption{Projected band structure (left) and projected density of states (right) on $s$ and $p$ orbitals of Si atoms for buckled silicene, each orbital is specified in the inset.}
    \label{fig:band_buckled}
\end{figure}

The band structure of buckled silicene shows some similarities to the graphene one. It is possible to see the Dirac states at the symmetric point K in the Brillouin zone crossing the Fermi level Figure \ref{fig:band_buckled}. The Dirac cone is a very interesting feature of the band structure dispersion. Near the Fermi level the states have a linear dispersion that can indicate a high electron mobility that can lead to several electronic applications in nanotechnology \cite{artigo_07}.

The projected band structure shows that the $p_z$ orbital is the one which contribute the most in the Dirac cones. It is possible to notice that their is a low contribution, thus relevant of $s$, $p_x$ and $p_y$ orbitals at the Fermi level. The system presents $sp^3$ hybridization. This result matches with the buckled stability explanation given before.  Silicene is usually called as zero gap semiconductor. With the inclusion of spin-orbit coupling, the energy gap of the buckled structure is in the order of 10$^{-3}$meV.

\begin{figure}[!htb]
\centering
\includegraphics[scale=0.02,clip,width=6cm]{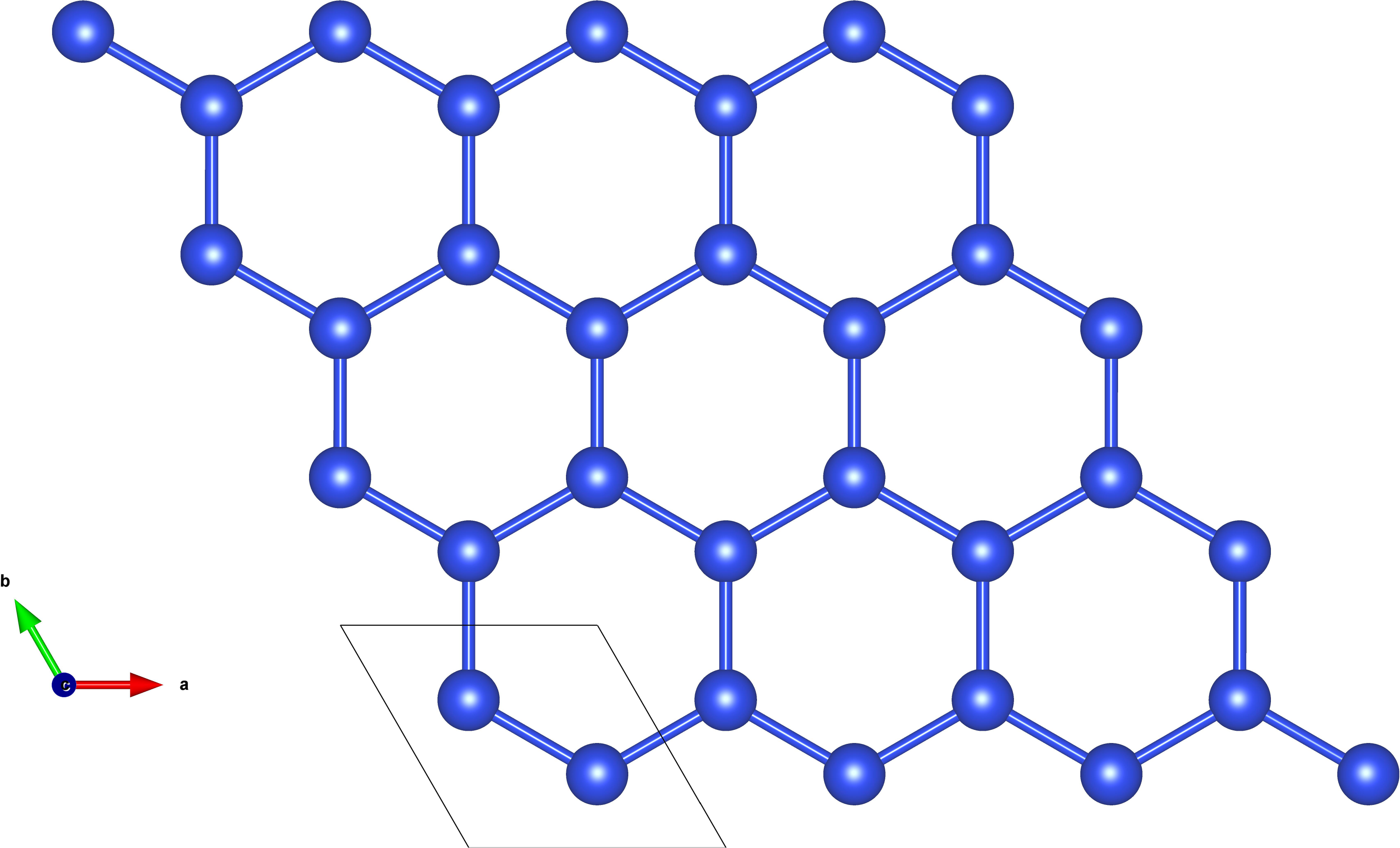}\\
\vspace{1cm}
\includegraphics[scale=0.02,clip,width=6cm]{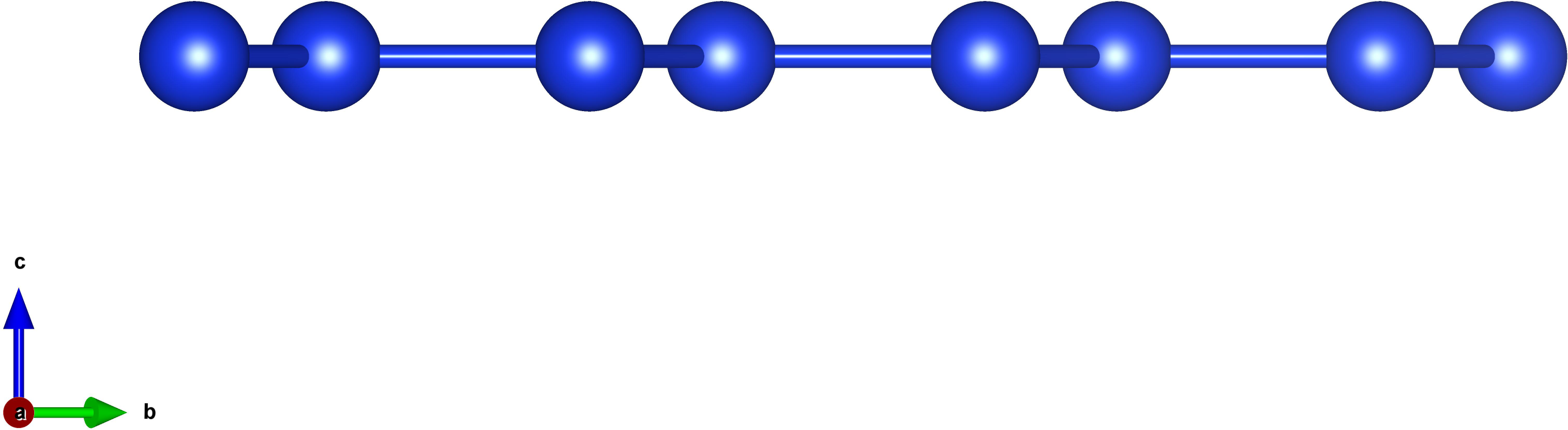}
\caption{a)Top and b) side views of planar monolayer silicene.}
\label{fig:siliceno_planar}
\end{figure}

Planar silicene has slightly higher energy and lattice parameter compared to buckled silicene, as expected. Despite of not being the ground state of the free-standing 2D silicon sheet is possible to obtain a local energy minimum for planar silicene, as seen in Figure\,\ref{fig:siliceno_planar}. 
We found a lattice parameter $a$ = 3.90\AA, the buckling distance is non existent and the bond length between silicon atoms is $d$ = 2.25\AA, in agreement with previous calculations\,\cite{PhysRevB.85.075423,artigo_33}.

\begin{figure}[!htb]
 \centering
     \includegraphics[scale=0.45,width = 8cm,clip]{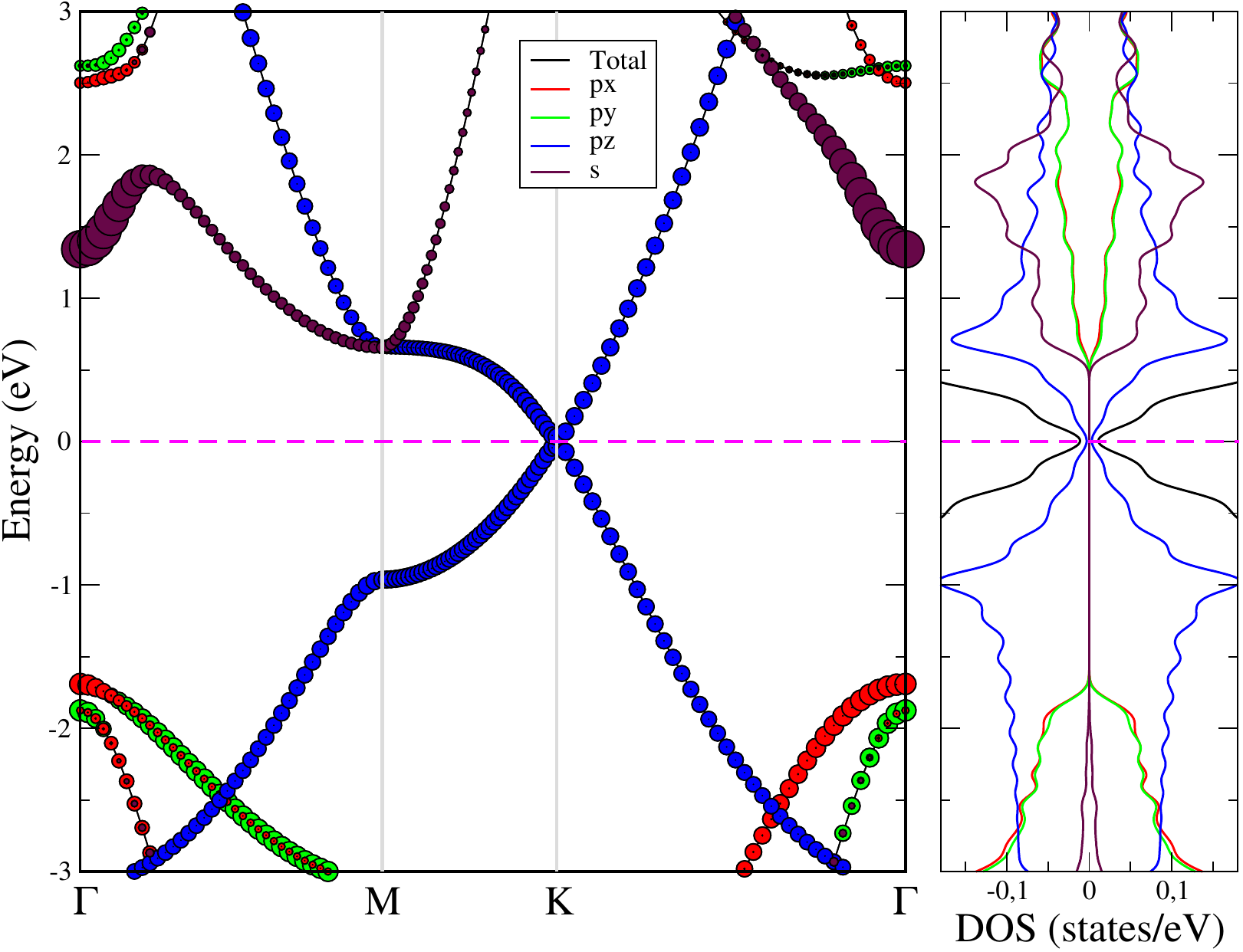}
    \caption{Projected band structure (left) and projected density of states (right) on $s$ and $p$ orbitals of Si atoms for planar silicene.}
    \label{fig:band_planar}
\end{figure}

The band structure  of planar silicene is very similar to  the buckled silicene band structure, one can notice that the Dirac cones  appear at the Fermi level at the K point. There is a small change in the density of states obviously because of the structural modification. The zero gap semiconductor characteristic still can be seen Figure\,\ref{fig:band_planar}.

\begin{figure}[!htb]
\centering
\includegraphics[scale=0.1,width = 6cm,clip]{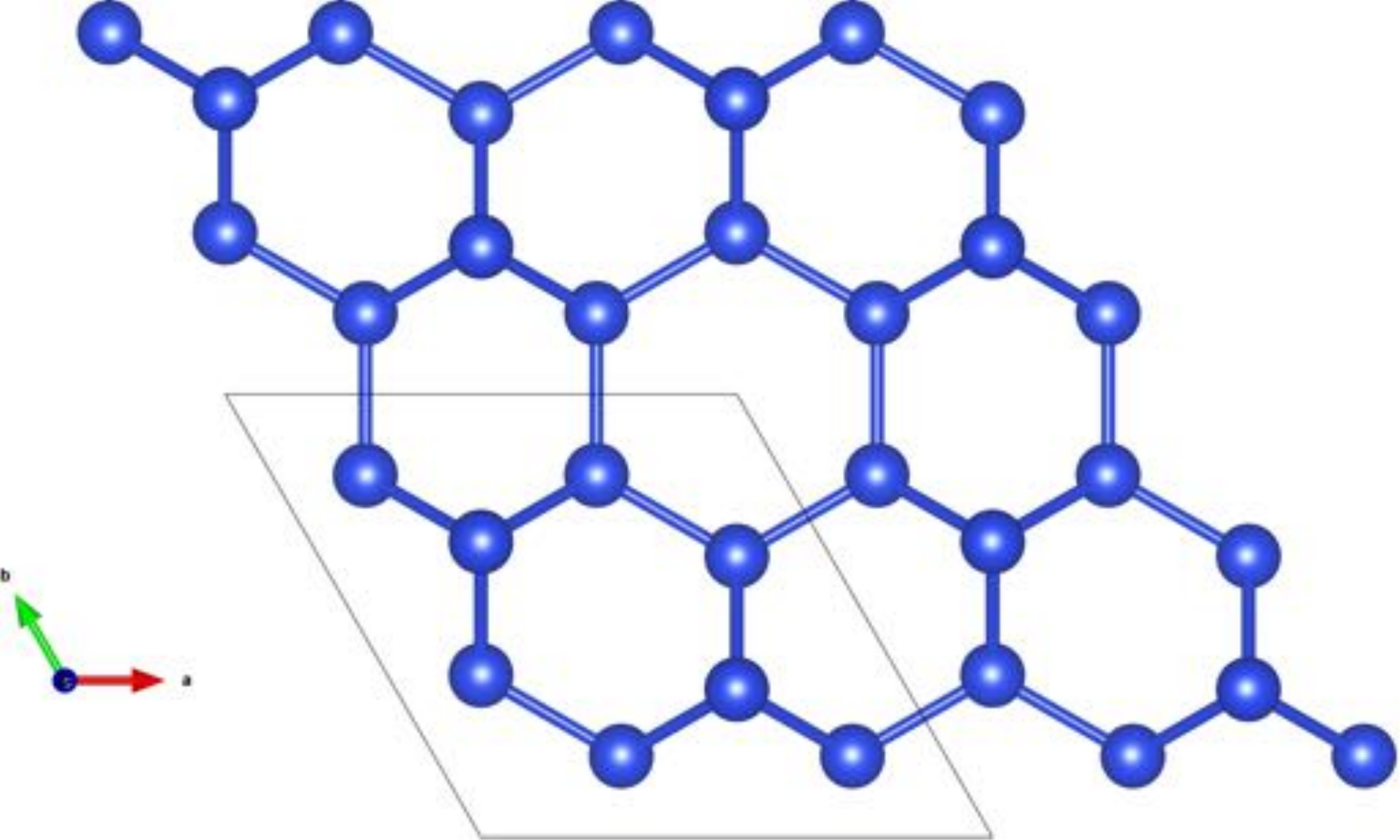}\\
\vspace{1cm}
\includegraphics[scale=0.02,width=6cm,clip]{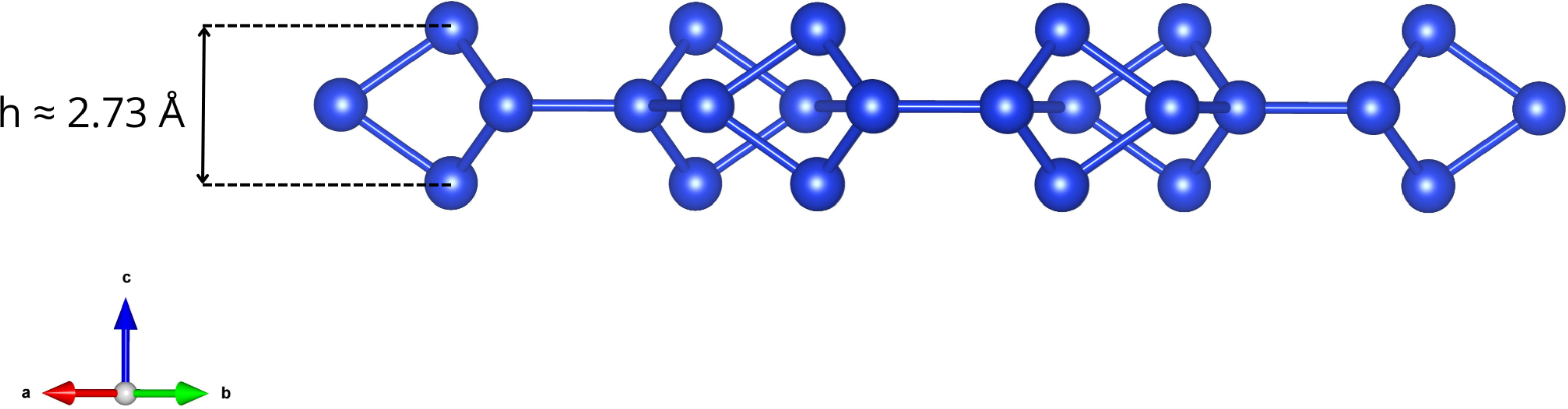}
\caption{Top and side views of  monolayer dumbbell silicene.}
\label{fig:dumbbell_geometry}
\end{figure}

\begin{figure}[!htb]
 \includegraphics[scale=0.45,width=8cm,clip]{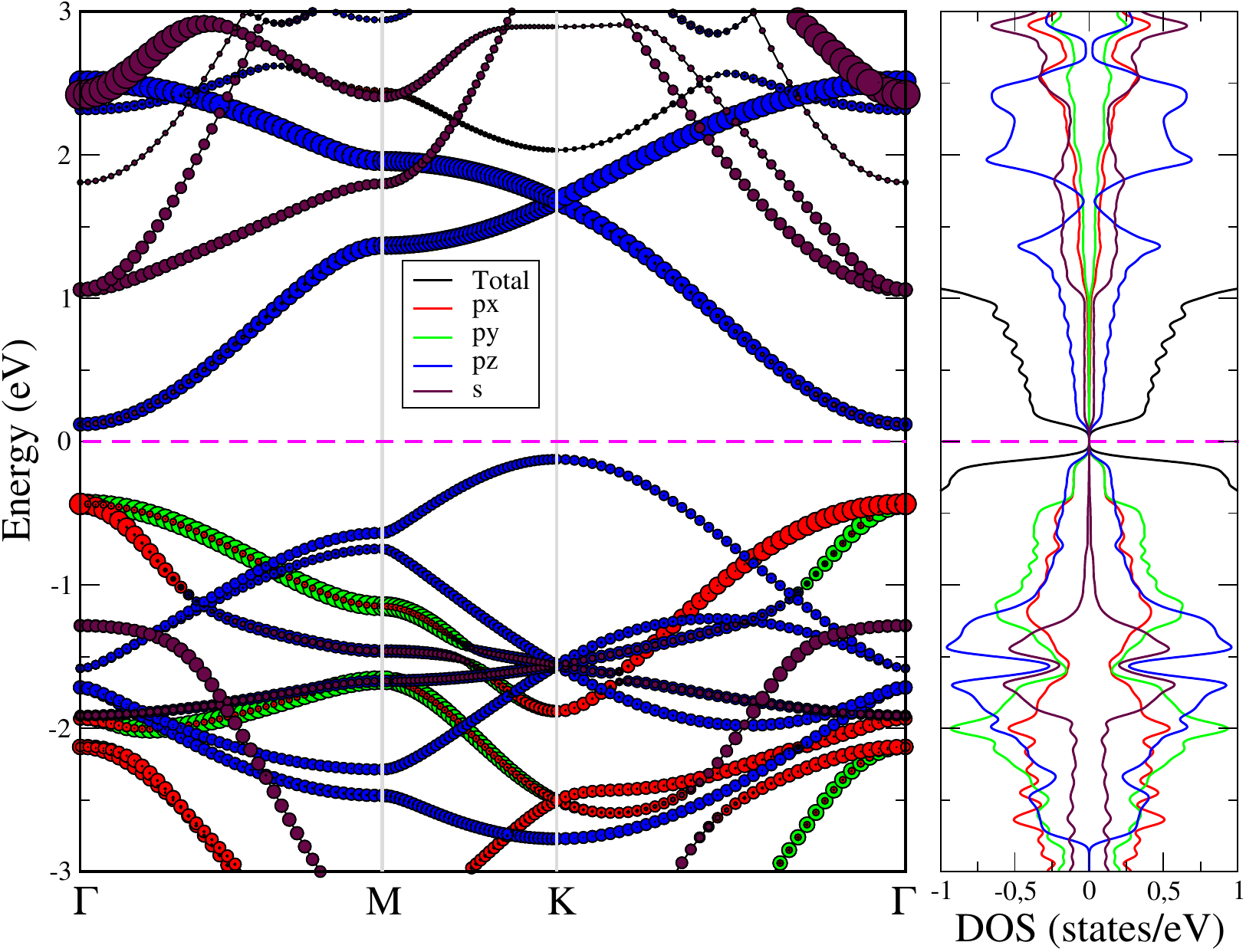}
\caption{Projected band structure (left) and projected density of states 
     (right) on $s$ and $p$ orbitals of Si atoms for dumbbell silicene.}
     \label{fig:dumbbell_band}
\end{figure}

Dumbbell silicene is a 2D silicon structure that derives from and adatoms adsorbed on buckled silicene as seen in Fig.\,\ref{fig:dumbbell_geometry}. Dumbbell silicene was recently synthesized on Ag(110), stabilized by adatoms\,\cite{artigo_29} has been demonstrated. This structure transformation leads to a band structure changes as shown in Figure \ref{fig:dumbbell_band}. Lattice parameter $a$ = 7.42\AA, the thickness is $d$ = 2.73\AA,  and the bond length between silicon atoms $d$ = 2.35-2.37\AA. The dumbbell projected band structure in Figure \ref{fig:dumbbell_band} shows us a gap formation of  0.25\,eV. The  valence band top is located at the K symmetry point in the Brillouin zone, and the bottom of conduction band is located at the $\Gamma$ point. One can notice that the orbital that contribute mostly at the Fermi level region is the $p_z$ orbital and in the conduction band.

\begin{table}[!htb]
\centering
\begin{tabular*}{8cm}{@{\extracolsep{\fill}}lcccc}
\hline
structure & $a$ (\AA) & $\Delta$ (\AA) & $d$ (\AA) & E$_{\rm g}$ (eV)\\ 
 \hline\hline
Bulk silicon & 2.73 & - & 3.36 & 0.46 \\ \hline
Buckled  & 3.87 & 0.38 & 2.27 & - \\ \hline
Planar  & 3.90 & 0.0 & 2.25 & - \\ \hline
Dumbbell  & 3.71 & 2.73 & 2.35-2.37 & 0.25 \\ \hline
\end{tabular*}
\caption{$\Delta$, $d$ and E$_{\rm g}$ for pristine bulk Si and silicene.}
\label{tab:stability}
\end{table}

The energy comparison of bulk silicon and silicene phases are shown in Figure \ref{fig:stability_silicene}. Their structural properties and electronic band gap  is shown in Table \ref{tab:stability}. We will use bulk silicon as reference to calculate the cohesive energy of the Si based 2D materials, and to calculate the cohesive energy of bulk silicon will be used the energy of silicon atoms set at free atoms. 

The cohesive energy to of the system is given by the equation below:

\begin{equation}
E_{\rm c} = E_{\rm silicene} - \sum_i \frac{E_{\rm ref}}{n_{\rm ref}}N_i
\label{eq3.1}
\end{equation}

\begin{table}[h!]
\begin{center}
\begin{tabular*}{8cm}{@{\extracolsep{\fill}}lcc}
\hline
structure & \multicolumn{2}{c}{E$_{\rm c}$(eV/atom)} \\
 \hline
  & ref. bulk  & ref. atom
 \\\hline
Bulk silicon & 0.0 & -4.58  \\ \hline
Buckled  & 0.64 & -3.94  \\ \hline
Planar  & 0.66 & -3.92  \\ \hline
Dumbbell  & 0.44 & -4.14  \\ \hline
\end{tabular*}
\caption{Cohesive energy E$_c$ using bulk silicon as reference (first column) and a free silicon atom as reference (second column).}
\label{tab:cohesive_energy}
\end{center}
\end{table}

It is possible to see that among the three two-dimensional Si-based structures, the one with lowest cohesive energy are dumbell and buckled silicene with very similar formation energy, as shown in Tab.\,\ref{tab:cohesive_energy}. A positive (negative) cohesive energy means an endothermic (exothermic) process.

\section{Chromium adsorption on silicene monolayers}

\begin{figure}[!htb]
    \centering
    \includegraphics[scale=0.8,clip,width=6cm]{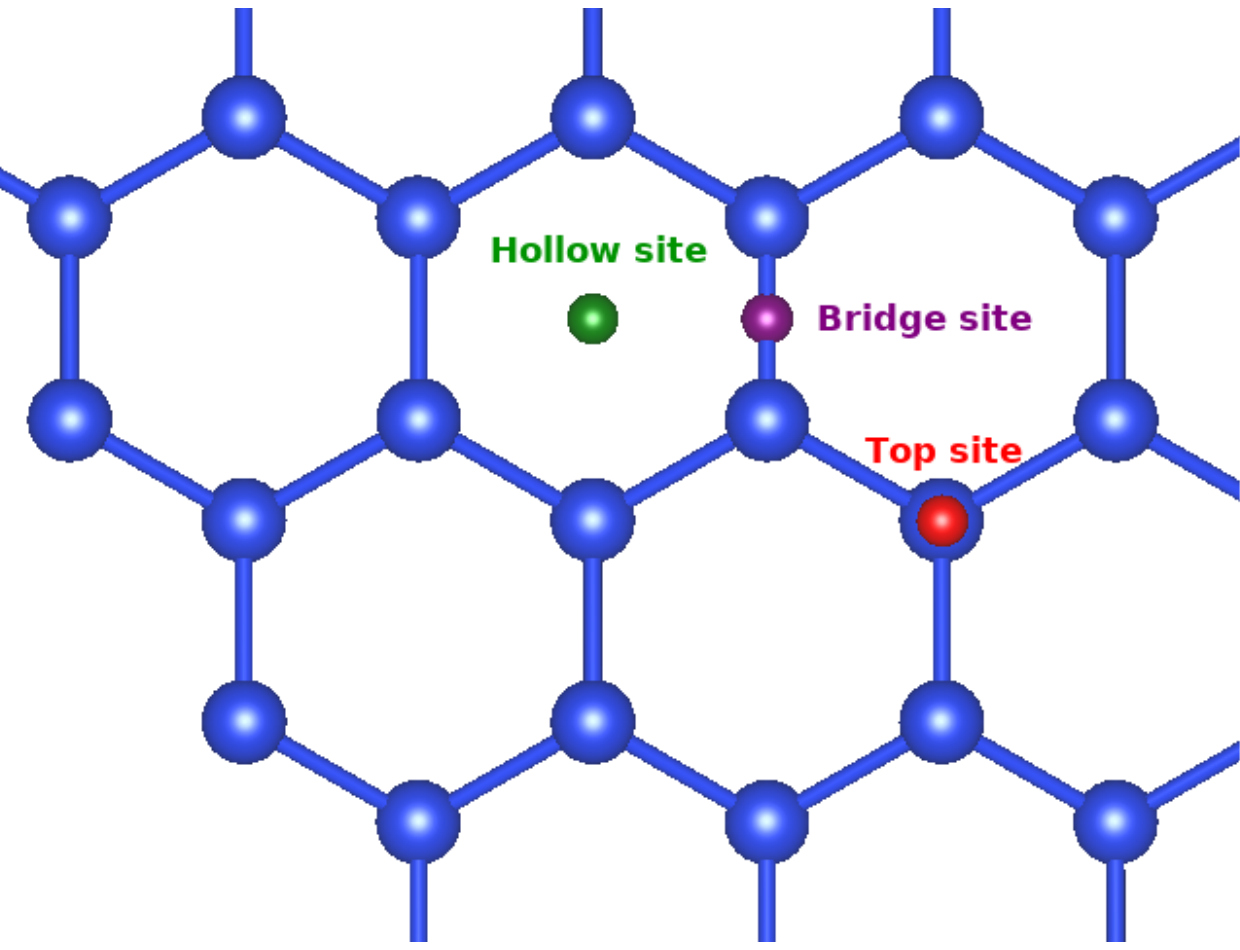}
    \caption{Adsorption sites for Cr atoms on silicene.}
    \label{fig:site_adsorption}
\end{figure}

The first we determine the most stable site adsorption site for the chromium atom. Three sites were investigated, represented in Figure \ref{fig:site_adsorption}: bridge (between two silicon atoms), hollow (in the middle of the hexagon) and top (on a single silicon atom). The follow configurations were set with a smart guess for their positions and then relaxations were performed to obtain the energy of each configuration.
 
The most stable configuration has a chromium atom adsorbed at a hollow site. The chromium atom does not stay on the bridge or  site, but relax  spontaneously to the hollow site

The lattice structural parametres are: $a$ = 4.08 {\AA},  the buckling distance $\Delta$ = 1.19{\AA}, and the bond length between silicon atoms is $d$ = 2.64{\AA} and magnetic moment of 3.14$\mu_B$.  The chromium  adsorption changes significantly the structure of silicene, affecting mainly the  buckling distance that goes from 0.45\AA  in the pristine monolayer to 1.19\AA in the adsorbed layer. 

This system has a peculiar characteristic related to the variation of
the lattice parameter. The ground state configuration is  where the Cr atom lies between the
silicon atoms of silicene. At this conformation the system is metallic
and shows a total magnetic moment of $3.14\mu_B$. If one applies a small strain on the unit cell,the Cr atom moves to the top of
the layer This
renders the silicene a semiconductor with a small gap and zero
magnetic moment. This is
an interesting finding since a small strain applied on the structure
induce a huge change making the material from metallic and magnetic to
a semiconductor with no magnetic moment. This result in in agreement
with predictions of \cite{Zheng2015,JAFARI2022}.

\begin{figure}[!htb]
    \centering
    \includegraphics[scale=0.45,width = 8cm]{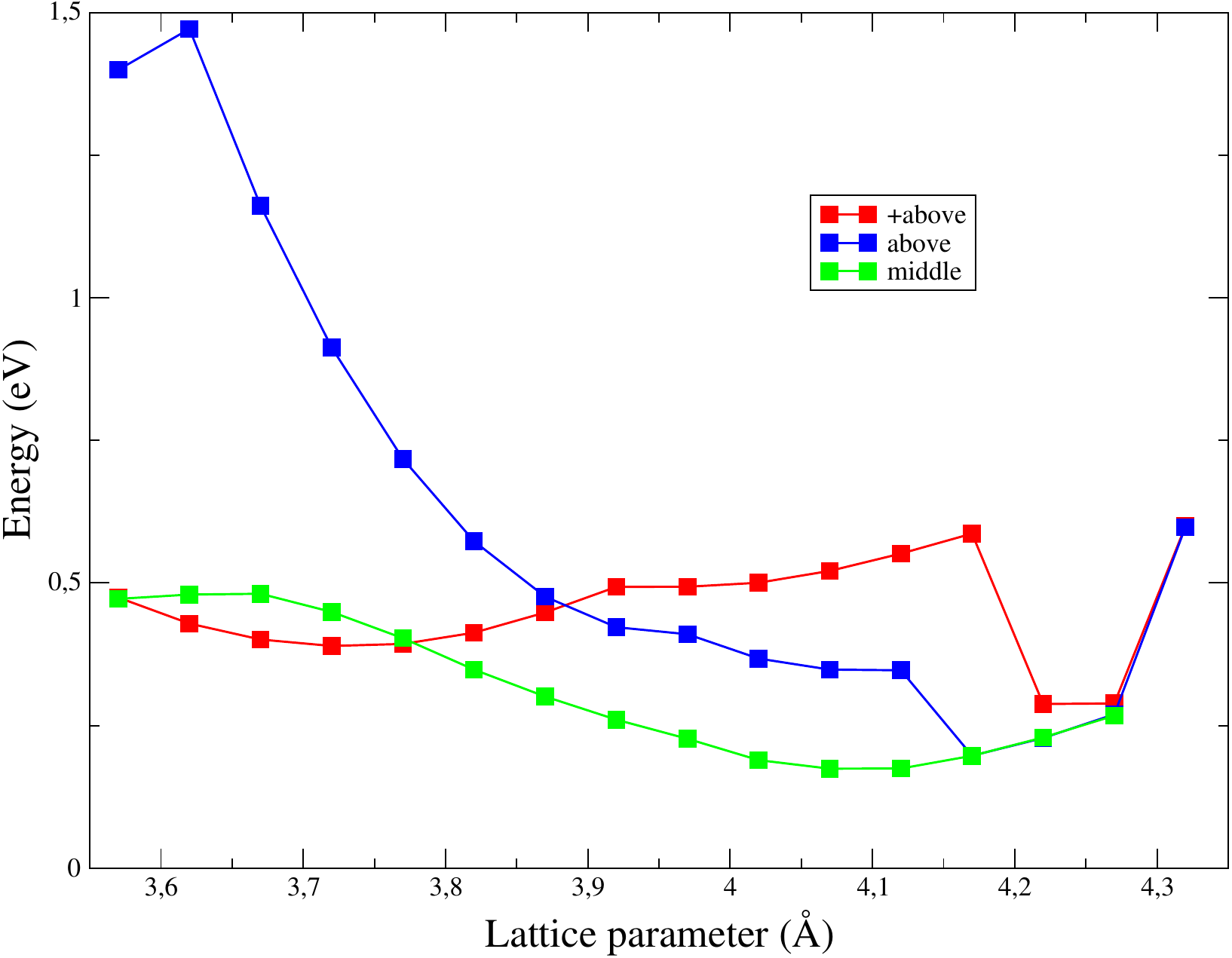}
    \caption{\label{fig:Cr_positions}Different heights of Cr atom position on buckled silicene.}   
\end{figure}

Figure \ref{fig:Cr_positions} shows the relative energy of the lattice constant for three different atomic heights of Cr on silicene.  The ground-state is very sensitive to the height, which could be useful to tune the magnetic properties os silicene. Furthermore, the Cr atom induces a strain on silicene, making the Si-Si distances  0.4 \AA longer, and the lattice parameter differ by 0.2\,\AA. 

\begin{figure}[!htb]
    \centering
    \includegraphics[scale=0.45,width = 8cm]{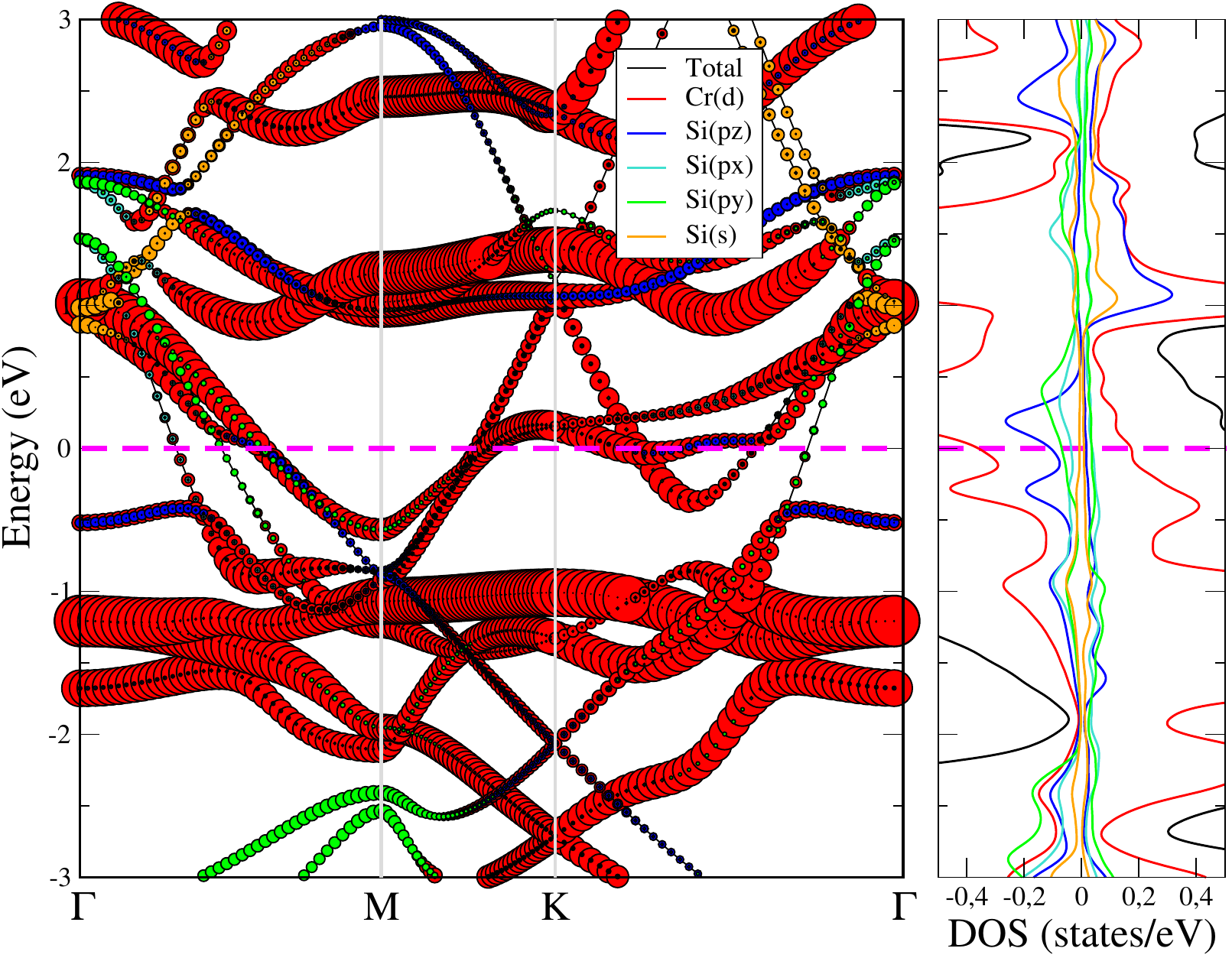}
    \caption{Projected band structure (left) and project density of state (right) for relaxed Cr-doped buckled silicene.}
    \label{fig:band_buckled_doped}
\end{figure}

The band structure of silicene doped with Cr atom shows that the systems goes from a zero gap semiconductor (pristine buckled silicene) with Dirac cones at K point, to a metal with a net magnetic moment, as seen in Figure \ref{fig:band_buckled_doped}. Those states at the Fermi level are mostly d orbitals from Cr atom. It is possible to notice that the $d$ orbitals from Cr atom hybridize with the $p_x, p_y, p_z$ orbitals from Si atoms, mostly with $p_z$. There is a negligible contribution from $s$ orbitals from Si atoms, which is the orbital that contribute mostly to the hybridization with $p_z$ on the pristine layer.  The Dirac cones are no longer at the Fermi level.

\vspace{6cm}

In order to identify better the hybridization between Cr atoms and Si atoms and the effect of quantum confinmene we compare the two forms 2D and 3D of CrSi$_2$. The bulk CrSi$_2$ \cite{artigo_30} has the structure shown in Figure \ref{fig:bulk_CrSi}, its has a hexagonal unit cell and is made of three Cr atoms and six nearest-neighbors Si atoms. The lattice parameter is $a=4.41\AA$ and $c=6.35\AA$. Bulk CrSi$_2$ shows a symmetric density of states, therefore is a non-magnetic material , and has a energy gap of E$_{\rm g}$ = 0.69eV, as one can see in Figure\,\ref{fig:band_buckled_doped}. 

The DOS of Cr doped buckled silicene Figure\,\ref{fig:band_buckled_doped} and DOS of bulk CrSi$_2$ shown in Figure\,\ref{fig:band_buckled_doped} it is possible to notice that the contributions of Si $p$ and $s$ to the hybridization with Cr $d$ orbitals are much larger in Cr doped buckled silicene. It means that the scattering of the Cr atoms by the Si atoms is much more intense in Cr doped buckled silicene than in bulk CrSi$_2$. It results in more localized states as one can see in Figure\,\ref{fig:chargedensity_diff}. More localized states results in better qubits that can be used in spintronics applications.

 The concentration of Cr atoms was varied from 2\% up to 50\%. The variation was made adding one Cr atom in the unit cell and then the size of the cell was increased. For example: for 50\% of concentration the unit cell used was the (1x1) silicene unit cell (2 Si atoms and 1 Cr atom), for 12,5\% of concentration the cell used was the (2x2) silicene cell (8 Si atoms 1 Cr atom).
 
 The initial position for each Cr atom on silicene is the same as calculated in the (1x1) unit cell. The Cr atom is positioned in the half height (define this) of silicene. Since  Cr atom produces a strain in the silicon honeycomb. It is important to notice that at low concentrations the lattice parameter does not change much compared to the pristine layer. Another aspect to be noticed is that the Cr atom increased the buckling distance of silicon sheet.

Adsorption energies of Cr atoms were  calculated with respect to bcc Cr (body-centered cubic) which is going to be our reference material for Cr atoms. Bcc chromium (two atoms in the atomic basis) is the most stable structure for pure chromium lattice with a lattice parameter of 2.85\AA. The metallic phase of chromium  has antiferromagnetic properties, all the Cr atoms presents magnetization and their total angular momentum are oriented anti-parallel in a way that the liquid magnetic moment of the crystal is zero. 

The reference for the silicon chemical potential is taken as the total energy for the pristine buckled silicene, for chromium the bcc phase. The general form of the energy adsorption equation is given below:

 \begin{equation}
     E_{\rm ads} = E_{\rm silicene/Cr} - \frac{E_{\rm silicene}}{N_{\rm silicene}}\cdot N_{Si} - \frac{E_{\rm bulk Cr}}{2}\cdot N_{\rm Cr}.
     \label{eq3.2}
 \end{equation}

\begin{figure}[!htb]
    \begin{center}
   \includegraphics[scale=0.35,width=7cm,clip]{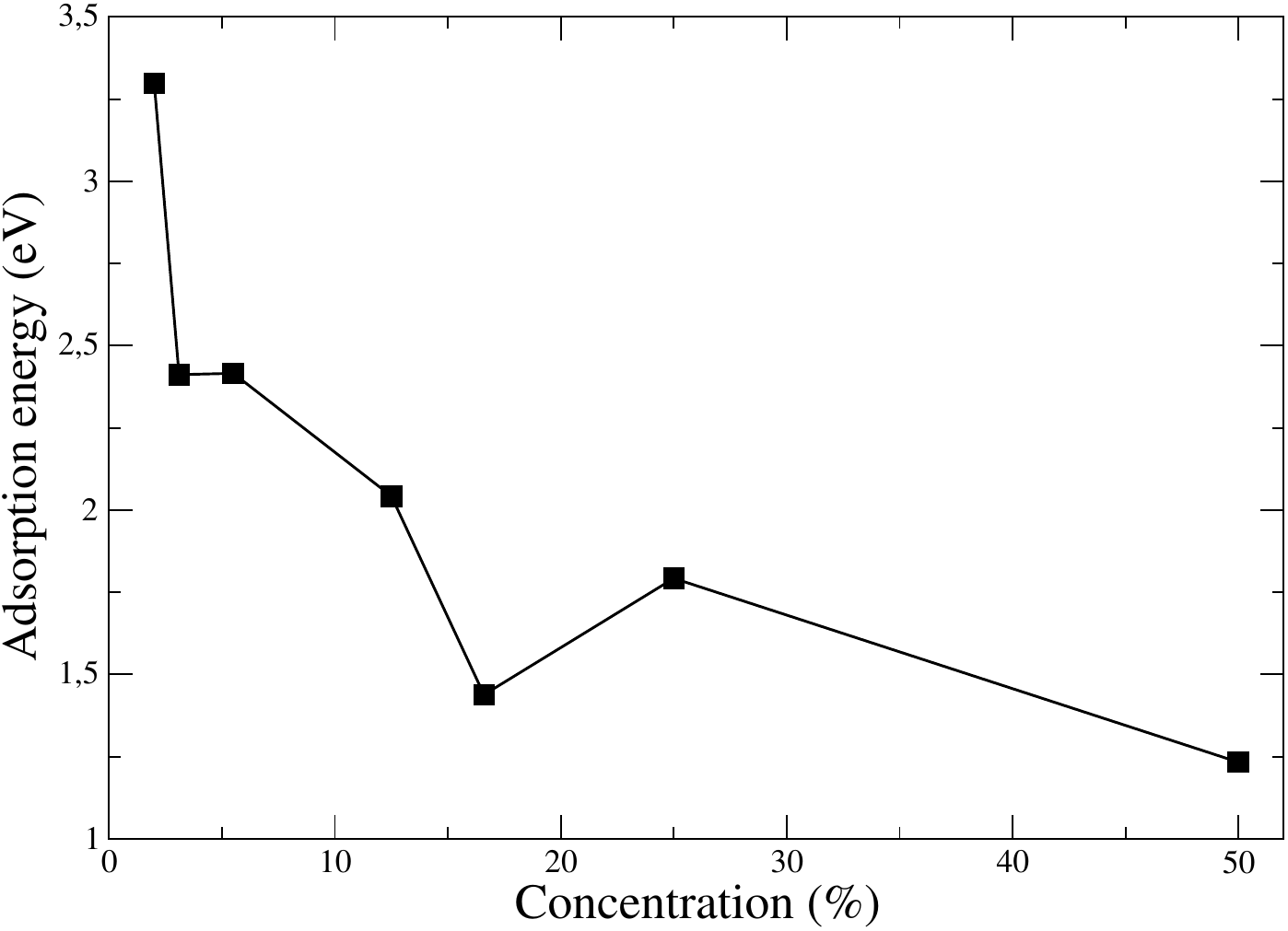}\\
     \includegraphics[scale=0.35, width = 7cm,clip]{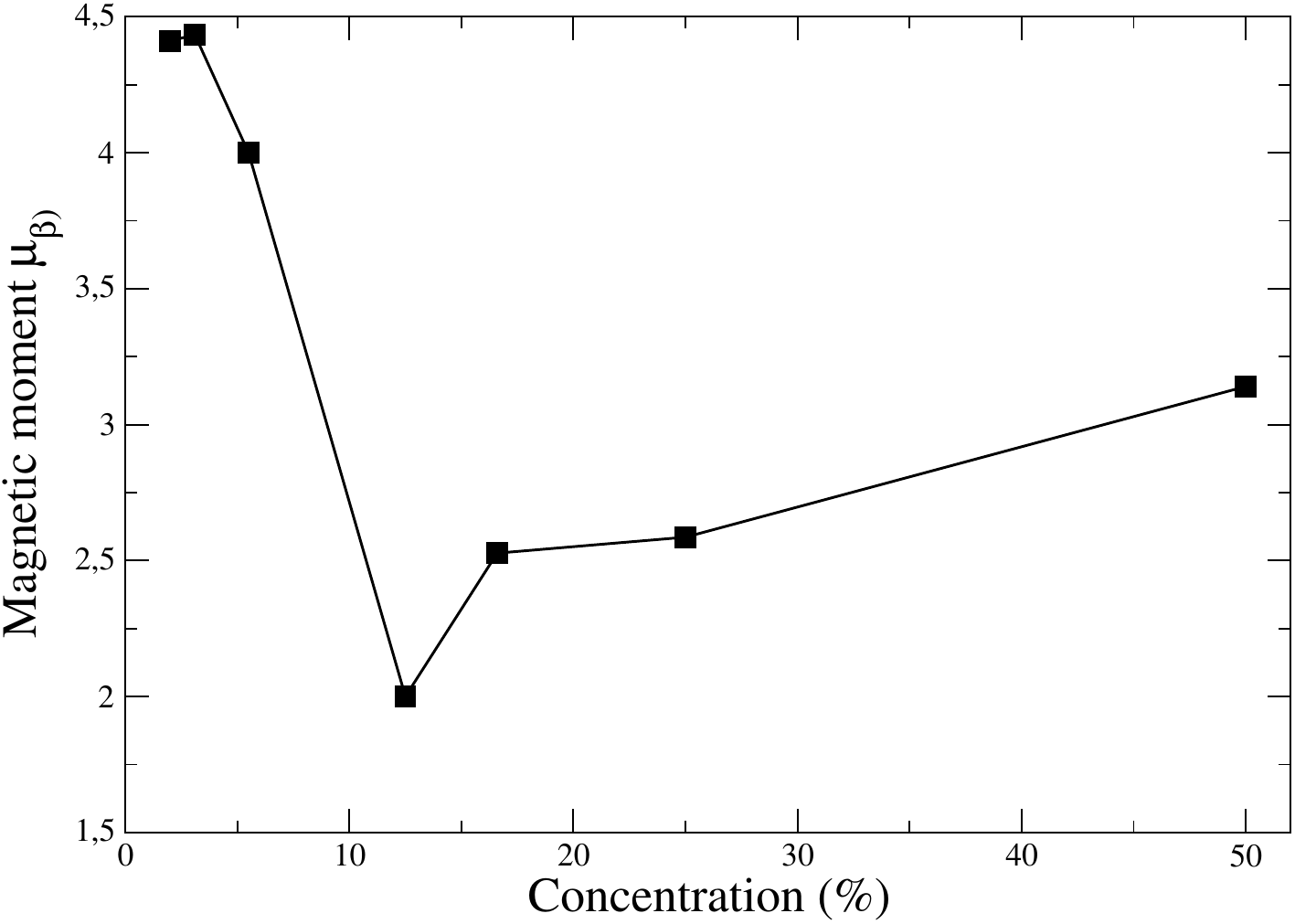}
     \end{center}
    \caption{Adsorption energy  and  magnetic moment of Cr atoms on silicene by concentration.}
    \label{fig:adsXconc}
\end{figure}

One can see that the adsorption energy decreases as the increase of concentration of Cr atom adsorbed in silicene. A possible explanation is the lateral interaction between Cr atoms, closer they are from each other, more they interact.  Another aspect to be analysed is the fact that all adsorption energies are positive.

Figure \ref{fig:adsXconc} b shows the magnetic moment as a function of the concentration. It is easy to notice that the magnetic moment decrease as the concentration of Cr atoms adsorbed increases.

\section{Substitutional Cr in silicene}

The substitution process is to remove a Si atom from the monolayer and adsorb in its place another type of atom. In this work the substitutional atom is the Cr atom and the formation energy of this kind of defect was calculated. The expression that calculates the formation energy is very similar to Equation \ref{eq3.2} and is based on Equation \ref{eq3.1}. The expression is:

\begin{equation}
    E_{\rm f} = E_{\rm silicene/Cr} - E_{\rm silicene} - E_{\rm chromium}
\end{equation}

The concentration of this kind of defect was varied and the formation energy was calculated.

\begin{figure}[!htb]
    \centering
    \includegraphics[scale=0.35,width = 7cm,clip]{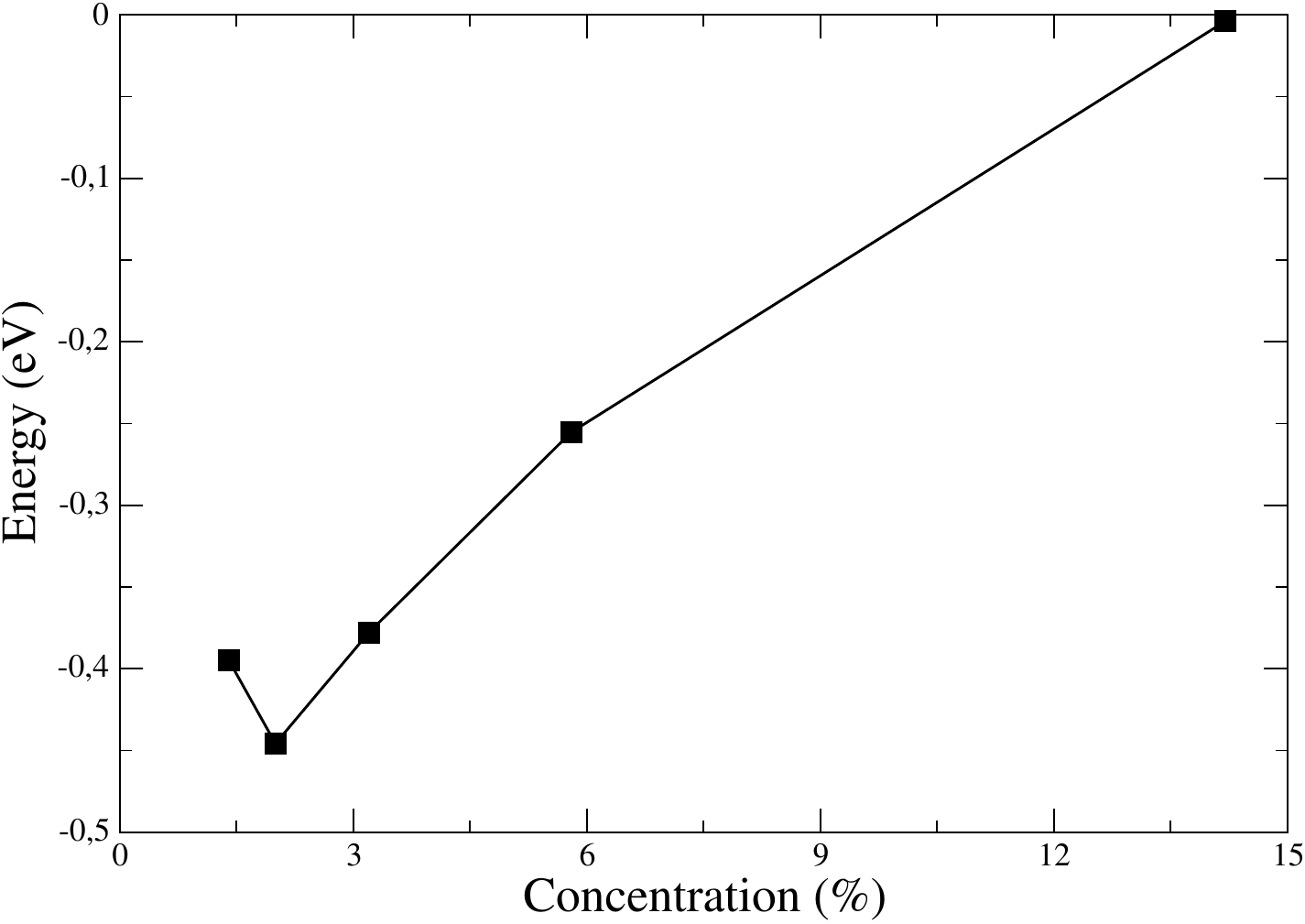}\\
     \includegraphics[scale=0.35,width = 7cm,clip]{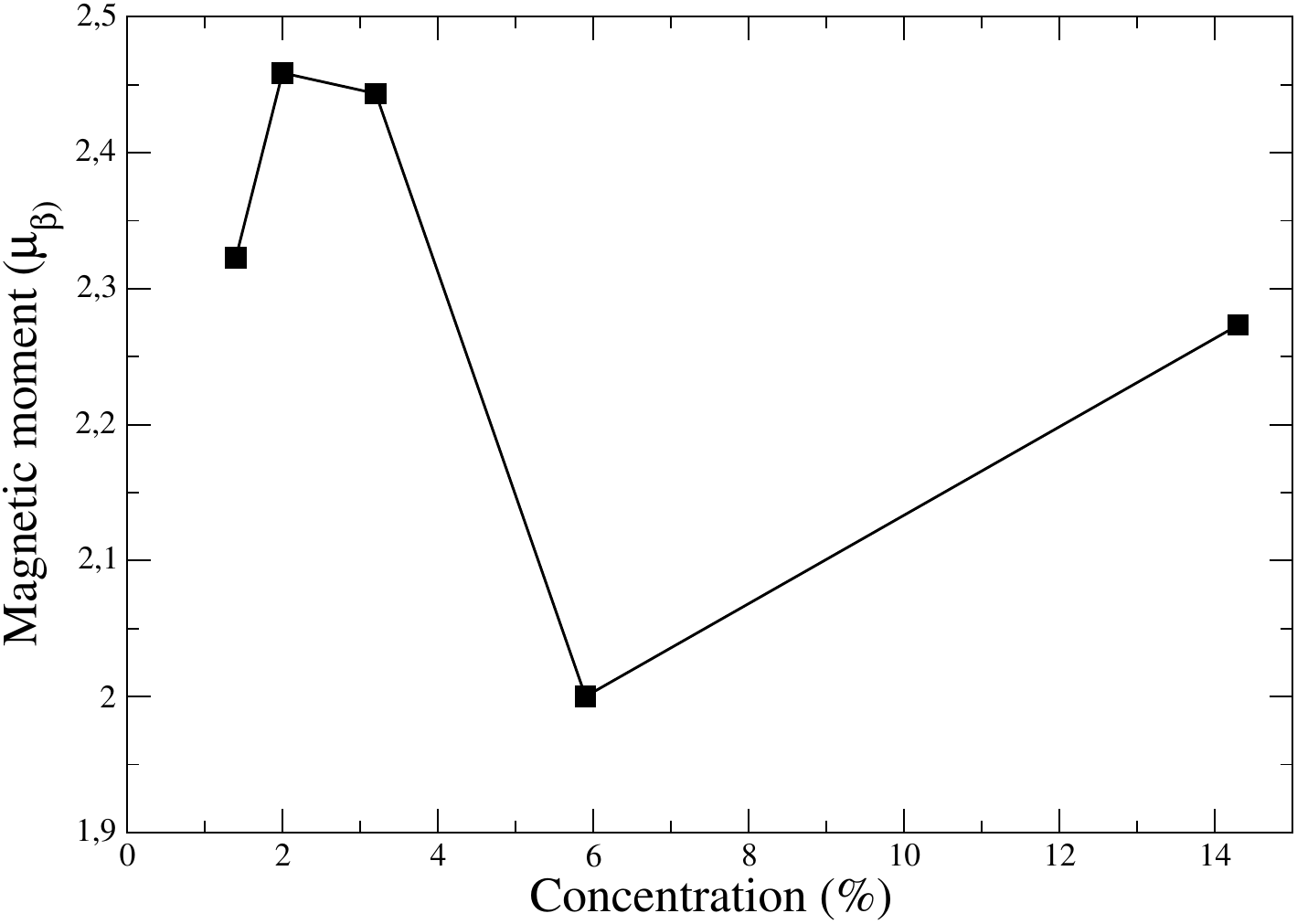}
    \caption{Formation energy and magnetic moment of substitutional Cr atom in silicene as a  function of the concentration.}
    \label{fig:form_energy}
\end{figure}

One can observe that all formation energies are negative, it means
that the system needs to lose energy to form the defect. For low
concentrations the formation energy is low. For higher concentrations
of the substitutional chromium, the formation energy is relatively
higher. It could be due to the Cr-Cr interaction that it becomes more
difficult to substitute Cr atoms when there is already a certain
number of substituted Cr atoms in silicene layer.

Figure \ref{fig:form_energy} shows the magnetic moment as a function
of substituted Cr atom concentration on silicene. After analysing the
graphic one can see that the magnetic moment is higher for lower
concentrations of Cr atoms. This finding is similar to adsorbed
chromium.

 The NEB method is used to determine the path which a atom on the crystal energy surface would take when connected by two states, which are minimum energy states of the system. NEB will seek for the less costly (in energy) path between two states. The path through the potential energy surface show us the energy barrier which the atom needs to overcome to go to a minimum energy state to another minimum energy state.

NEB method was performed to calculate the energy barriers for the Cr atom on the potential energy surface of silicene mono layer. The NEB was performed in two cases: Cr atom goes from a hollow site to an adjacent one, along a path parallel or perpendicular to silicene plane, as shown in Figure \ref{fig:barrier_geometry}. 

\begin{figure}[!htb]
    \centering
\includegraphics[scale=0.3, width = 6cm,clip]{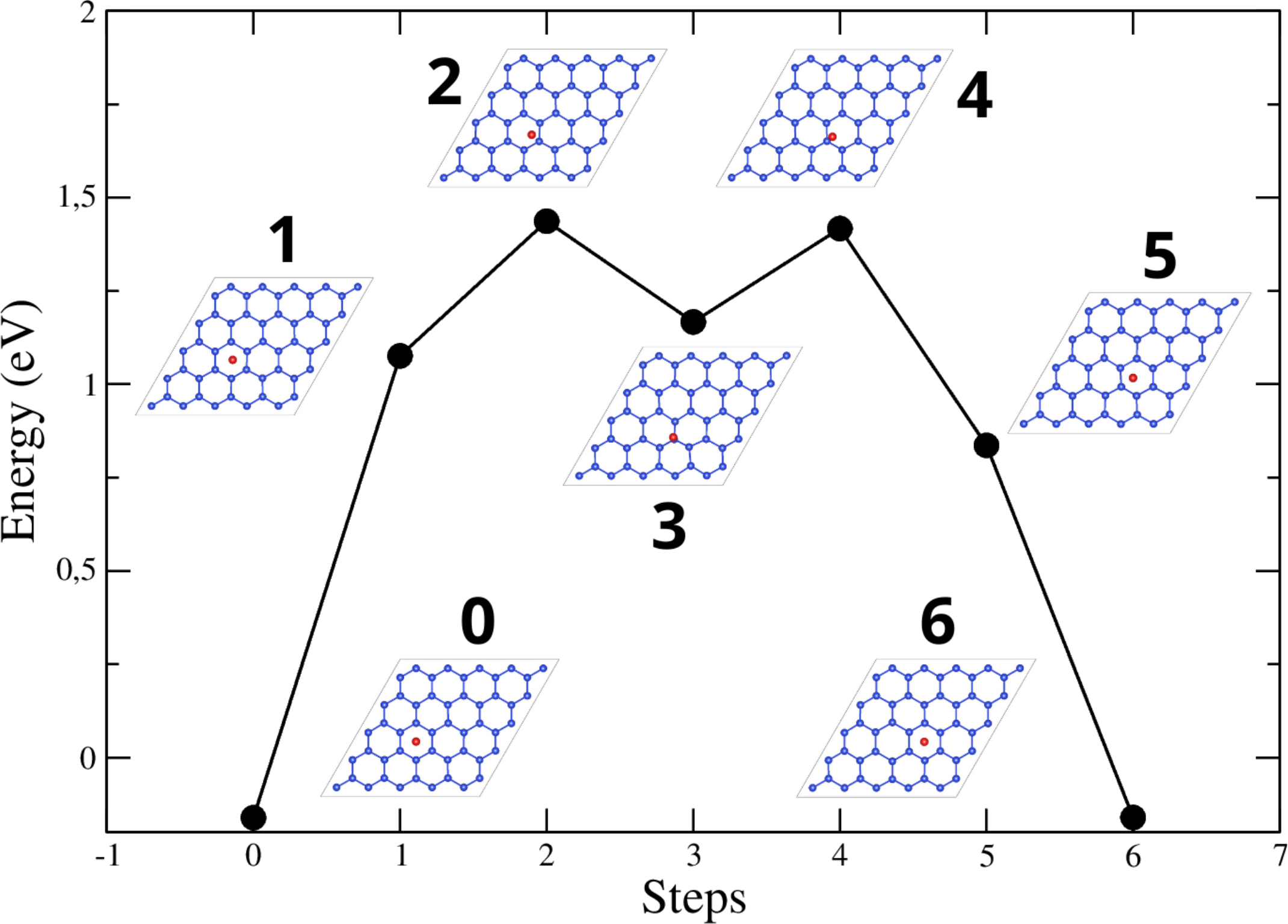}\\
  \includegraphics[scale=0.5, width = 6cm,clip]{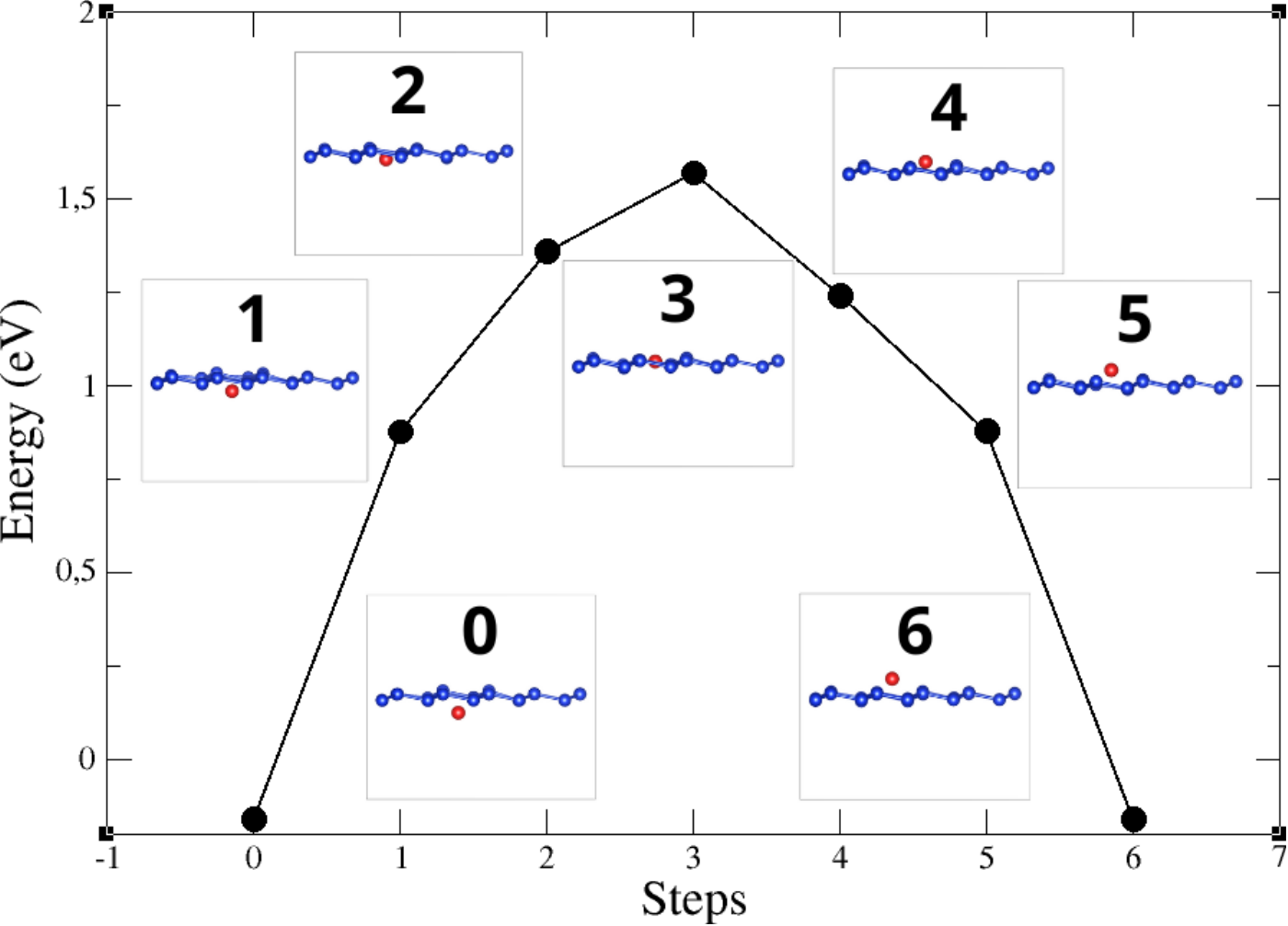}
    \caption{Diffusion barrier for parallel and perpendicular path.}
    \label{fig:barrier}
\end{figure}

The diffusion barrier energy for the parallel path is 1.60\,eV. The
barrier has a significantly high energy, which means that it is
unlikely to a Cr atom goes from one hollow site to an adjacent
one. The barrier presents a local minimum energy state in the top of
the barrier, represented by the step 3 showed in Figure
\ref{fig:barrier}. This shape is due to the path, which passes by a
minimum local energy site which is a top site. Top site is a high
symmetry point in the hexagonal silicene lattice, usually high
symmetry sites are lower energy states which explains the formation of
local minimum energy point. This shape of barrier could be associated
with buckled 2D structures.  The radius of the atom influences the
adsorption position. Thus, a question is whether it is possible to
find a minimum local energy pit (as step 3 in Figure \ref{fig:barrier}
of a certain 2D buckled material adsorbed with a certain atom that
could be captured and maintained adsorbed in the top site instead of
the global minimum point. The diffusion barrier for the path
perpendicular to the layer is 1.73eV. The energy of this barrier is
higher compared to the path parallel to the layer. The step 3 in
Figure \ref{fig:barrier} shows the transition states is where the Cr
atom passes through the monolayer. This point can be related to the
maximum strain that the Cr atom is putting on the lattice of silicene.

\begin{figure*}[!htb]
\centering
  \includegraphics[scale=0.04,width=7cm, clip]{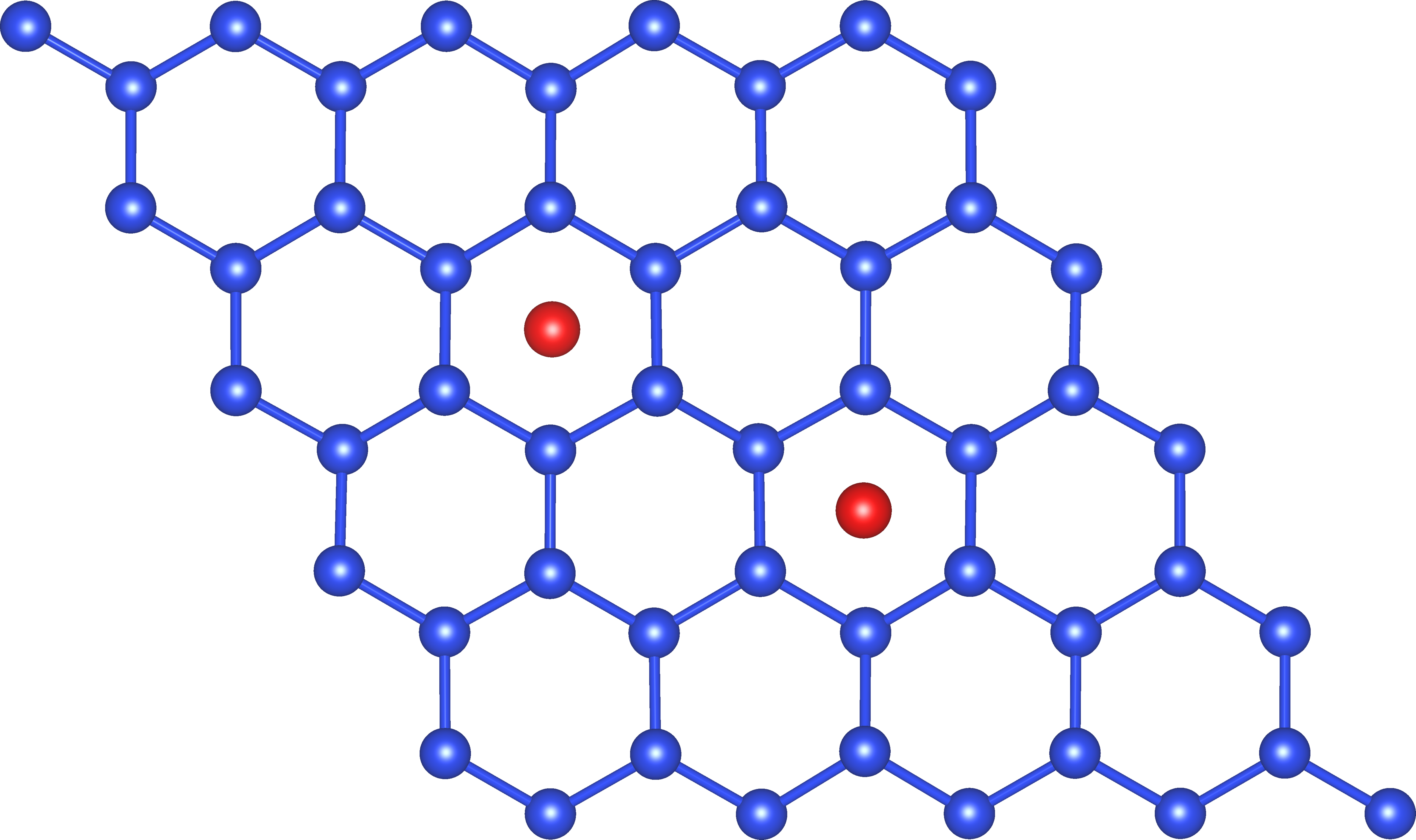}
   \includegraphics[scale=0.04,width=7cm, clip]{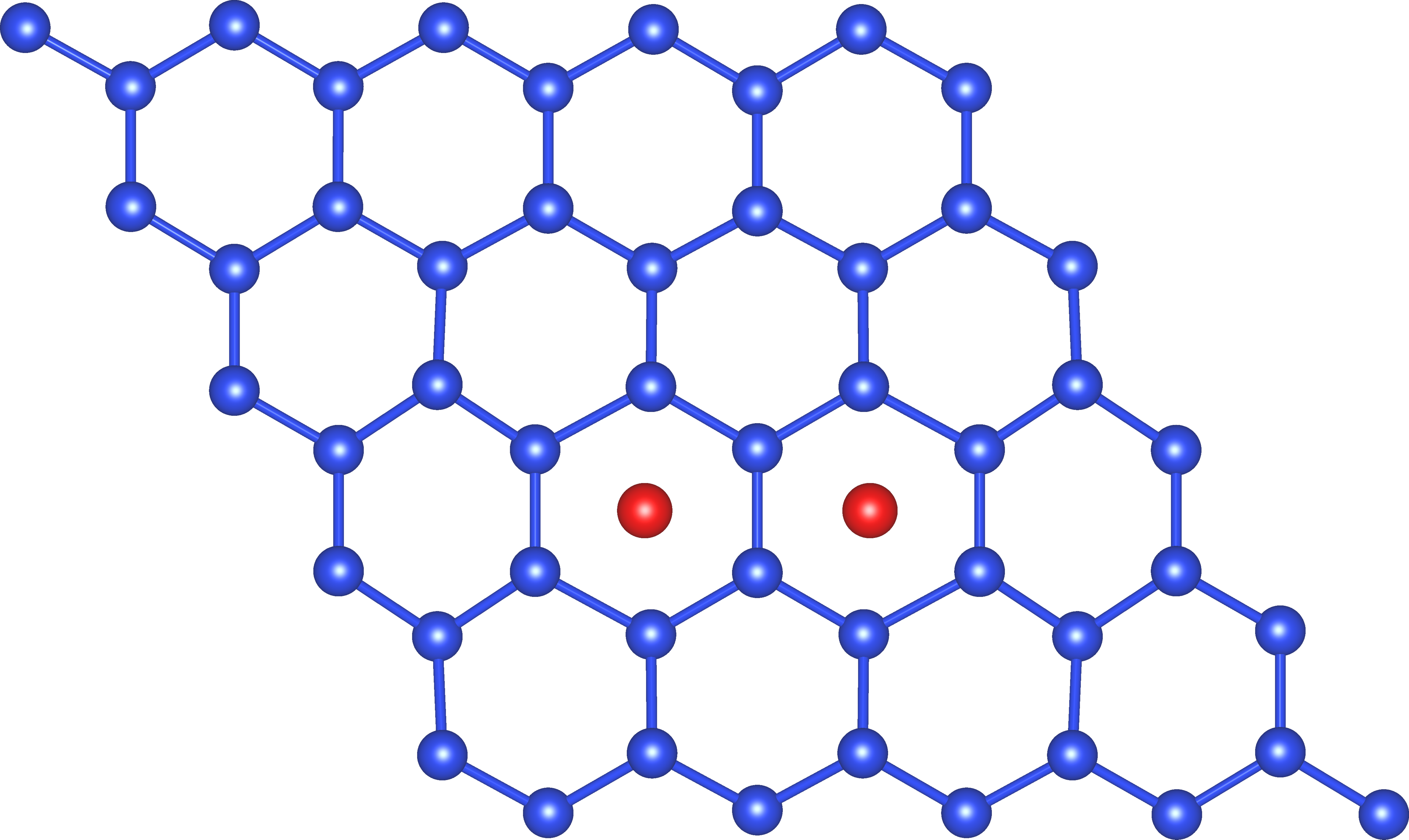}
    \caption{Chromium atom pair adsorbed on silicene along the a) arm-chair direction and b) zig-zag direction.}
    \label{fig:Crpair_direction}
\end{figure*}

\begin{table*}[!htb]
\centering
\begin{tabular*}{12cm}{@{\extracolsep{\fill}}lccccccc}
\hline
 & \multicolumn{1}{c}{$\Delta$ E (eV)} & \multicolumn{2}{c}{a({\AA})} & \multicolumn{2}{c}{$\mu_{\rm tot}(\mu_B)$} & \multicolumn{2}{c}{d$_{\rm Cr-Cr}$ (\AA)} 
 \\ \hline
conc.(\%)    & & AFM & FM & AFM & FM & AFM & FM \\ \hline
5.5 & -0.16 & 3.90 & 3.92 & 0.00 & 3.61 & 4.16 & 4.18 \\ \hline
3.1 & -0.26 & 3.90 & 3.92 & 0.00 & 4.26 & 4.19 & 4.19 \\ \hline
2.0 & -0.33 & 3.90 & 3.92 & 0.00 & 4.70 & 4.20 & 4.15 \\ \hline
1.3 & -0.47 & 3.90 & 3.92 & 0.00 & 4.53 & 4.22 & 4.20 \\ \hline
\end{tabular*}\caption{Energy difference between AFM and FM states $\Delta$E, lattice parameter a, magnetic moment $\mu_{\rm tot}$ and  Cr-Cr distance d$_{\rm Cr-Cr}$ for silicene monolayer doped with a pair of Cr atom along the zigzag direction.}
\label{tab:energy_diff_zigzag}
\end{table*}

\begin{table*}[!htb]
\centering
\begin{tabular*}{12cm}{@{\extracolsep{\fill}}lccccccc}
\hline
         & \multicolumn{1}{c}{$\Delta$ E (eV)} & \multicolumn{2}{c}{a(\AA)} &  \multicolumn{2}{c}{$\mu_{\rm tot}(\mu_B)$}  & \multicolumn{2}{c}{d$_{\rm Cr-Cr}$ (\AA)}\\ 
 \hline
conc.(\%) & & AFM & FM & AFM & FM & AFM & FM \\ \hline
5.5 & 0.09 & 3.87 & 3.88 & 0.00 & 3.61 & 7.04 & 7.07 \\ \hline
3.1 & 0.10 & \multicolumn{1}{c}{3.87} & 3.88 & \multicolumn{1}{c}{0.00} & 4.26 & \multicolumn{1}{c}{7.06} & 7.04 \\ \hline
2.0 & -1.44 & \multicolumn{1}{c}{3.87} & 3.88 & \multicolumn{1}{c}{0.00} & 4.70 & \multicolumn{1}{c}{7.03} & 6.83 \\ \hline
1.3 & -1.51 & \multicolumn{1}{c}{3.87} & 3.88 & \multicolumn{1}{c}{0.00} & 4.53 & \multicolumn{1}{c}{7.04} & 6.77 \\ \hline
\end{tabular*}
\caption{Energy difference between AFM and FM states $\Delta$E, lattice parameter a, magnetic moment $\mu_{\rm tot}$ and Cr-Cr distance d$_{\rm Cr-Cr}$ for silicene monolayer doped with Cr-pair  along the arm-chair direction.}
\label{tab:energy_diff_armchair}
\end{table*}

\begin{table}[!htb]
\centering
\begin{tabular*}{8cm}{@{\extracolsep{\fill}}lcc}
\hline
 &  \multicolumn{2}{c}{$\Delta E$ (eV)}   \\ \hline
conc. (\%) & AFM & FM \\ \hline
5.5 & -0.92819 & -1.18623 \\ \hline
3.1 & -0.98085 & -1.3428  \\ \hline
2.0 & -1.01236 & 0.10065  \\ \hline
1.3 & -1.01284 & 0.03537  \\ \hline
\end{tabular*}
\caption{Energy comparison between arm-chair and zig-zag directions in antiferromagnetic and ferromagnetic states.}
\label{tab:energy_diff_directions}
\end{table}

The magnetic properties of the chromium atom  on silicon monolayer has been investigated for silicene with a pair of Cr adsorbed along two different directions (arm-chair and zig-zag) as shown in Figure \ref{fig:Crpair_direction}.   The energy difference between AFM and FM for both directions is shown in Tables \ref{tab:energy_diff_armchair} and \ref{tab:energy_diff_zigzag}. 

\subsection{Chromium atom-pairs adsorbed on dumbbell silicene}

\begin{figure*}[!htb]
\centering
\includegraphics[scale=0.045,width = 5cm,clip]{supercell.pdf}
\includegraphics[scale=0.15,width = 4cm,clip]{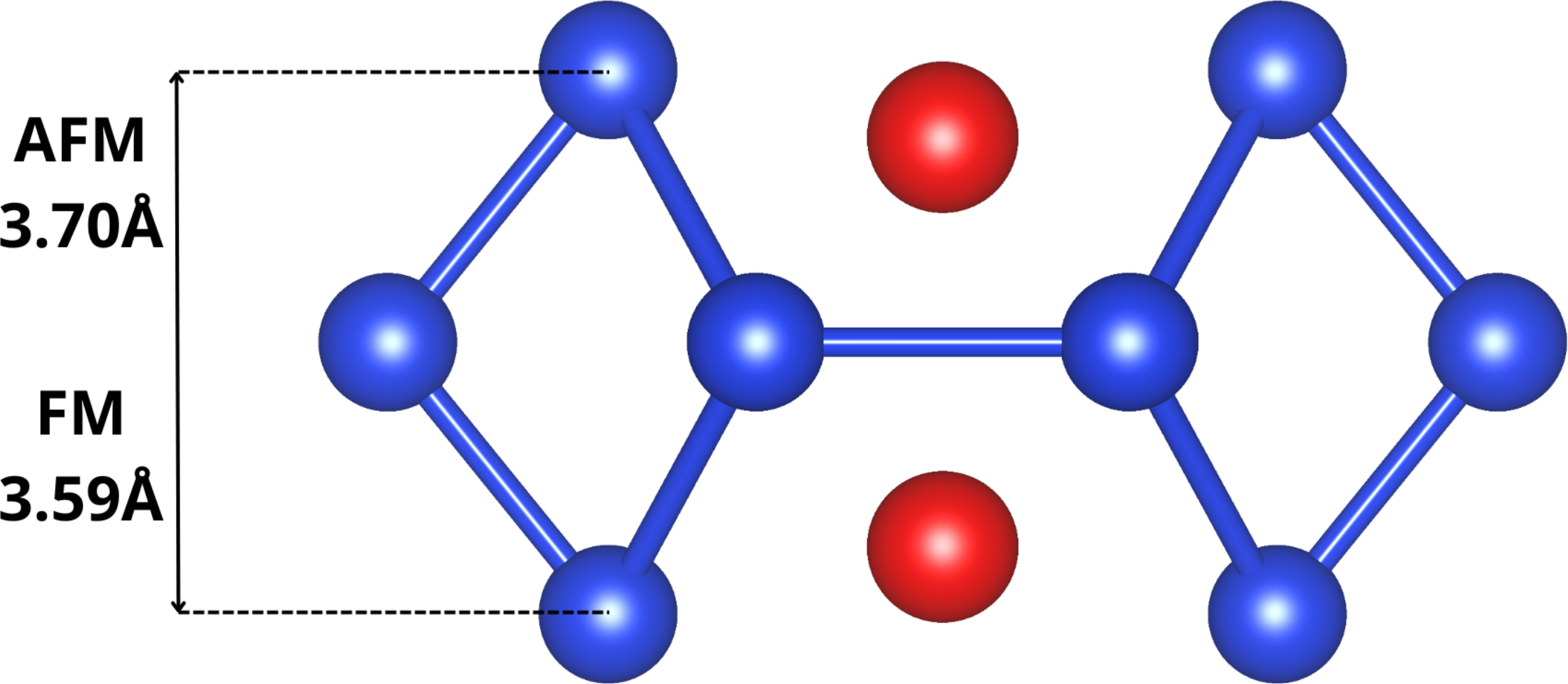}
\includegraphics[scale=0.045,width = 3cm,clip]{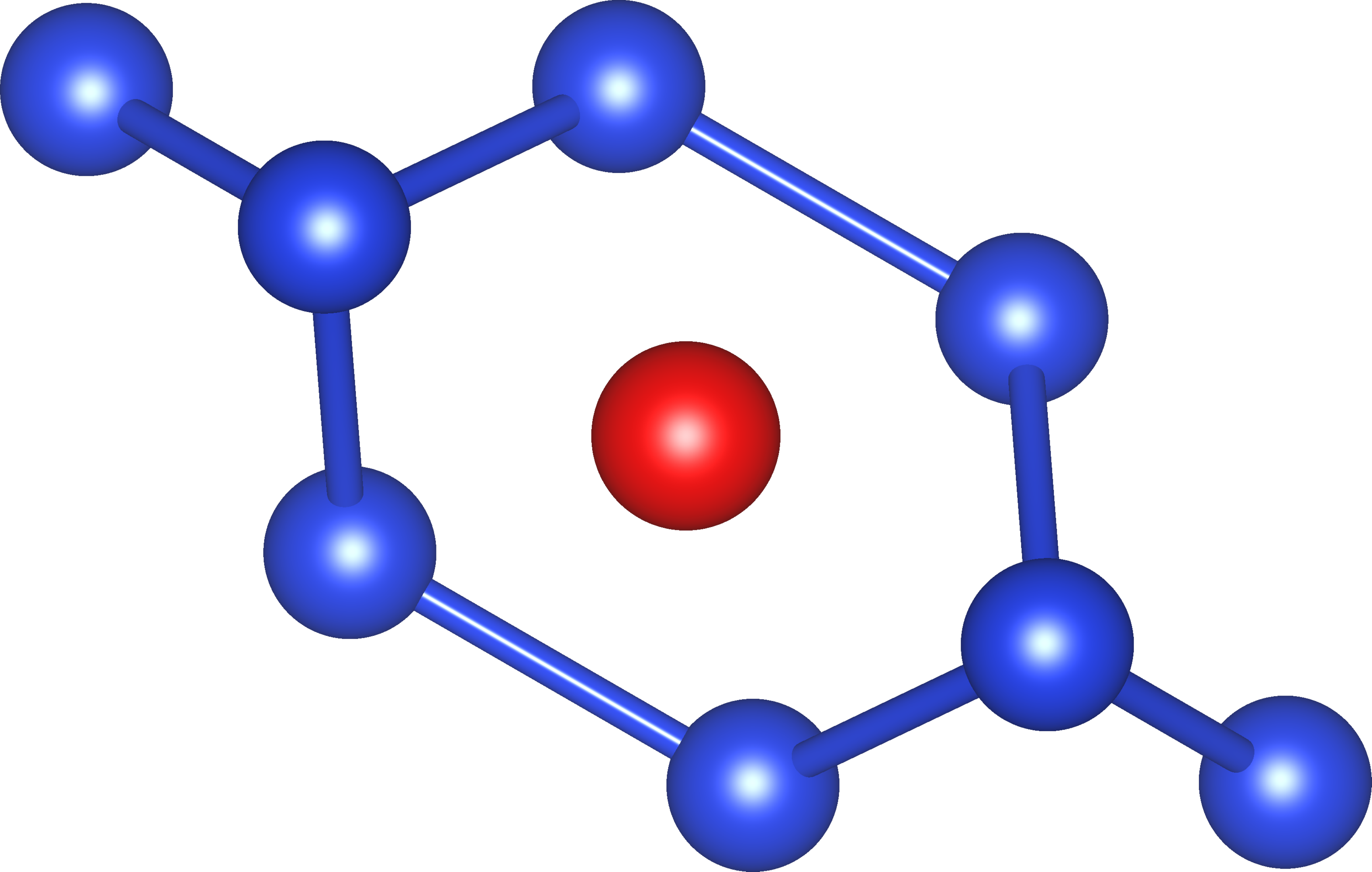}
    \caption{a)Side view b) top view  of dumbbell silicene unit cell doped with two Cr atoms.}
    \label{fig:dumbbell_sideview}
\end{figure*}

The pristine dumbbell silicene was doped with two Cr atoms as shown in Figure \ref{fig:dumbbell_sideview}. A feature to be noticed is that the height of dumbbell silicene increased around 1 {\AA}  and the lattice parameter decreased slightly when doped with Cr atoms.Cr atoms tends to shrink the dumbbell silicene in the layers plane and increase its height Table \ref{tab:dumbbell_pair}. By comparing the properties of AFM and FM states of DB silicene it is notable that FM state presents more magnetic moment per Cr ion in FM state. This property might be associated with the disposition of the Cr atoms in the DB silicene structure since in AFM state has their pair of Cr atoms closer to each other  at a distance of $2.79\AA$ compared to FM state where the Cr-Cr distance is $3.35\AA$. However, the AFM state is more stable than FM state by 0.6\,eV.

\begin{table}[!htb]
\centering
\begin{tabular*}{8cm}{@{\extracolsep{\fill}}lcccc}
\hline
 & a (\AA) & $\mu$/Cr ($\mu_B$) & h & d$_{\rm Cr-Cr}$ (\AA) \\ 
  \hline
AFM &
  7.12 &
  1.72 &
  3.70 &
  2.79 \\ \hline
FM &
  7.12 &
  2.69 &
  3.59 &
  3.35 \\ \hline
\end{tabular*}
\caption{$a$, $\mu$/Cr, $h$ and d$_{\rm Cr-Cr}$ for both AFM and FM states in dumbbell silicene  upon adsorption of pair of Cr atoms.}
\label{tab:dumbbell_pair}
\end{table}

\begin{figure}[!htb]
    \centering
  \includegraphics[scale=0.45, width=7cm]{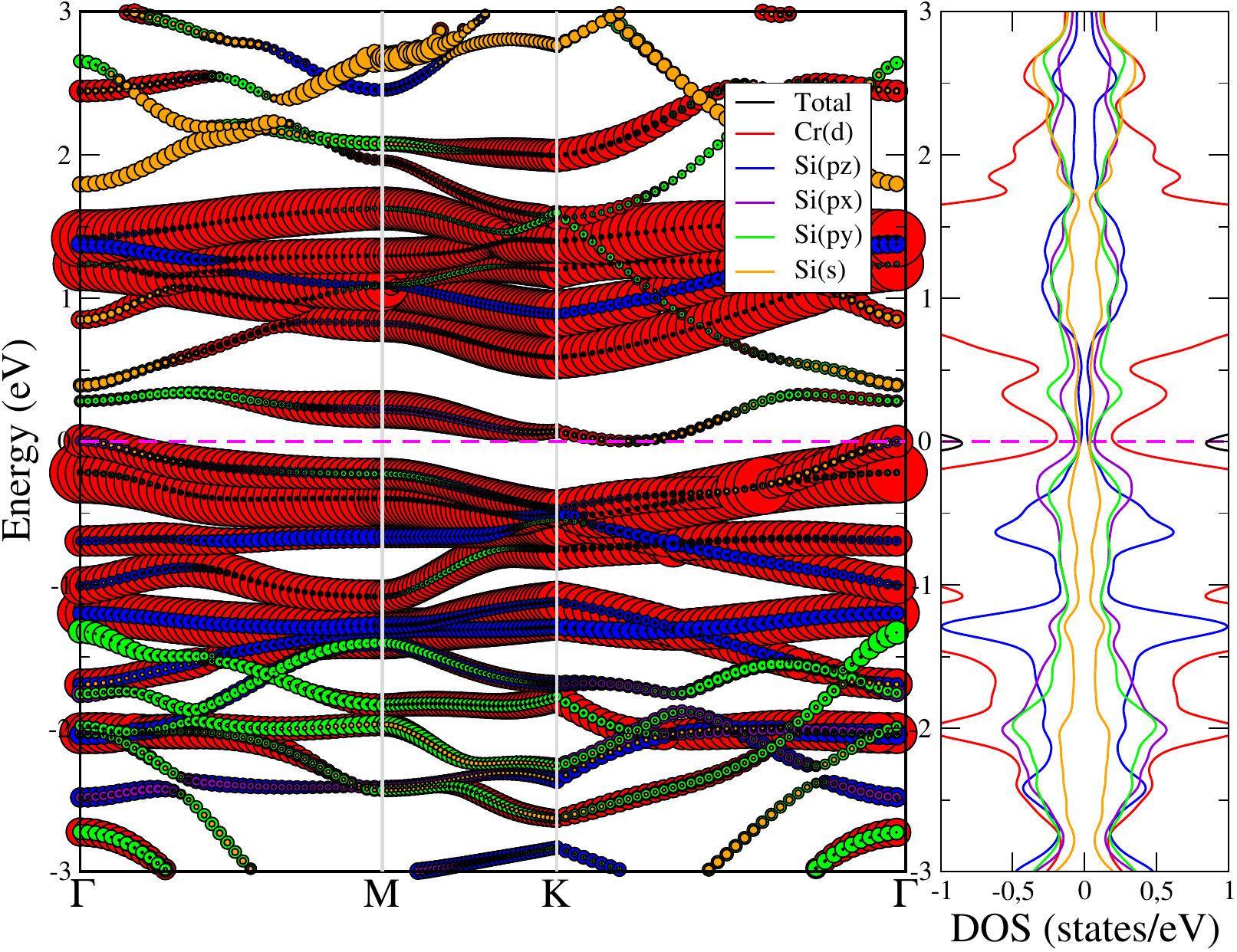}\\
   \includegraphics[scale=0.45, width=7cm]{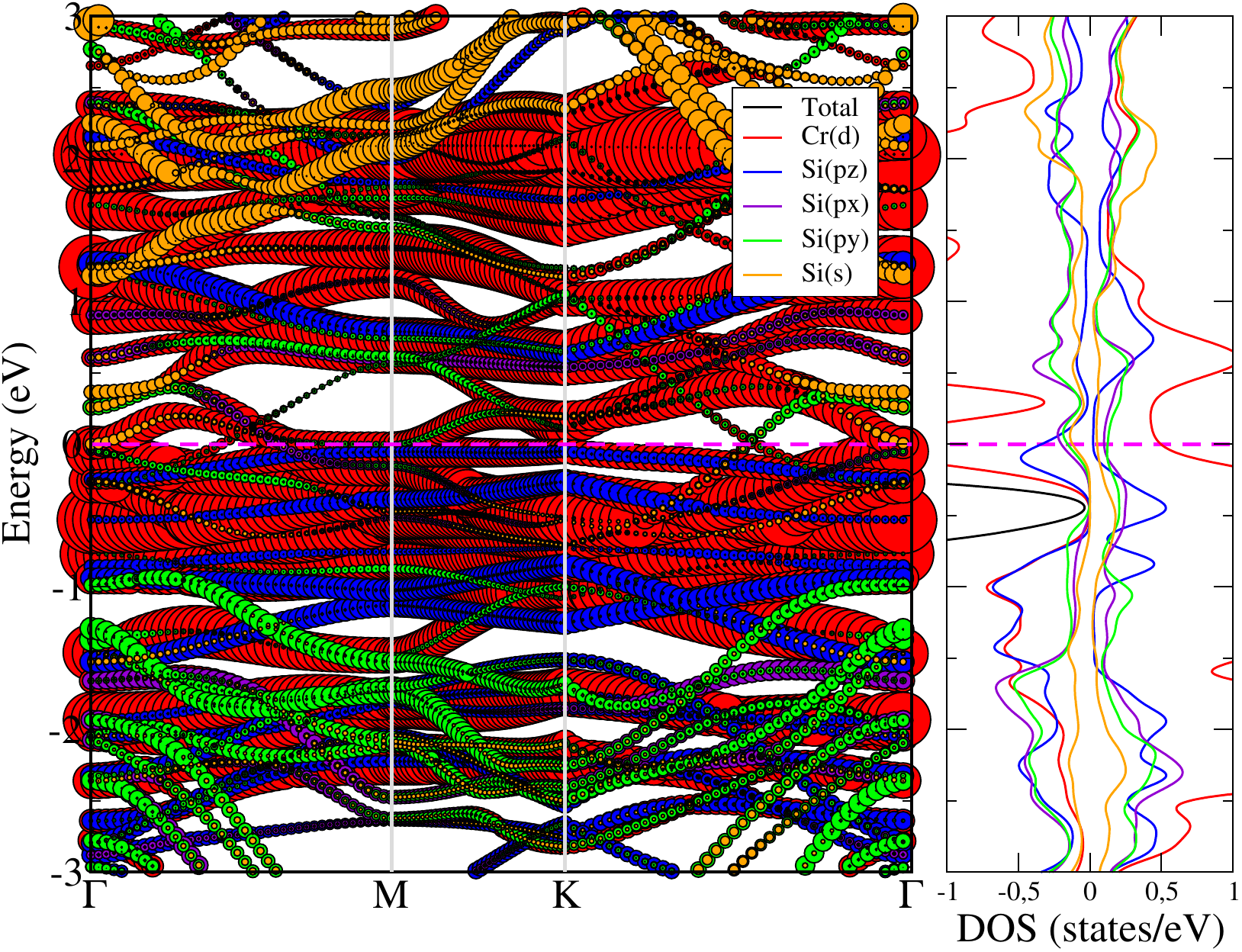}
    \caption{Projected density of state and projected band structure of dumbbell silicene for a) AFM and b) FM configurations.}
    \label{fig:band_afm_fm}
\end{figure}

As silicene dumbbell is doped with Cr atoms, the structure becomes metallic. It is possible to notice the difference between AFM and FM density of states (DOS). The contributions of Si orbitals at the Fermi level of AFM state is very small Figure \ref{fig:band_afm_fm}. Near the Fermi level the orbitals that contribute the most are $p_x$ and $p_y$. The contributions of Si orbital at the Fermi level of FM state are slightly larger compared to AFM state. Near the Fermi level it is possible to notice a significant larger contribution of $p_z$ orbital in FM DOS, seen in Figure \ref{fig:band_afm_fm}. This larger contribution of Si $p_z$ orbital and smaller contribution of Si $p_x$ and $p_y$ orbitals to the hybridization with Cr $d$ orbitals might be related to the increase of the magnetic moment of structures studied in this work.
 
 \section{Pristine bilayer silicene}

Bilayer silicene can be obtained on  a (111) Ag surface\,\cite{ARAFUNE2013,artigo_27}. Few theoretical reports have shown that the preferred stacking is the AB\,\cite{artigo_33,KHARADI2023,Fu2014,Do2019}.
One can stack a monolayer right above the other, this kind of stacking called AA stacking. One can stack the two monolayers in a way that the lattice points of one layer is shifted, those are known as AB stacking. These two type of stacking can be branched because of the buckling pattern. It results in the six bilayer structures that have been studied in this work.

\begin{figure*}[!htb]
     \centering
  \includegraphics[scale=0.4,width=12cm,clip]{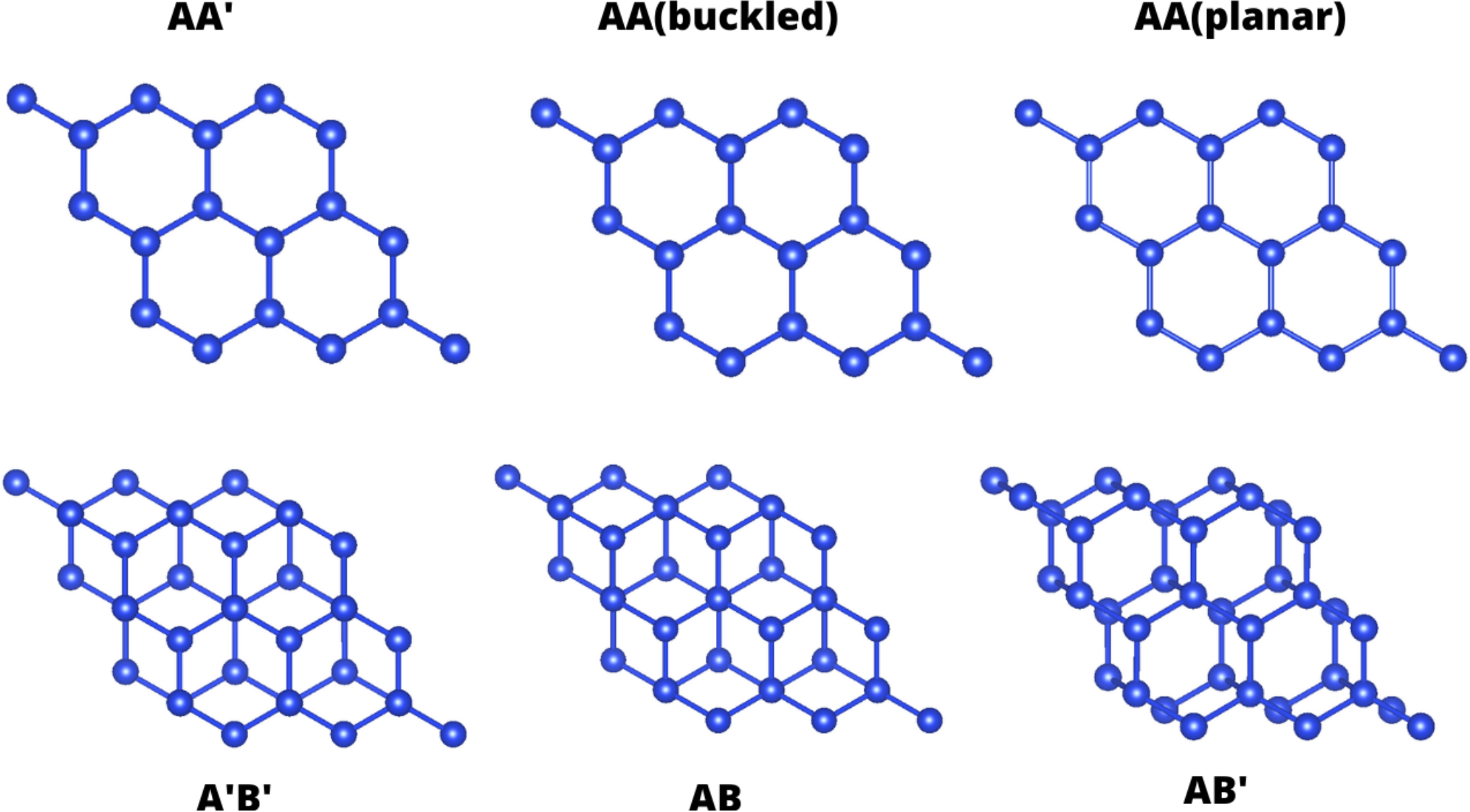}\\
    \includegraphics[scale=0.4,width = 12cm,clip]{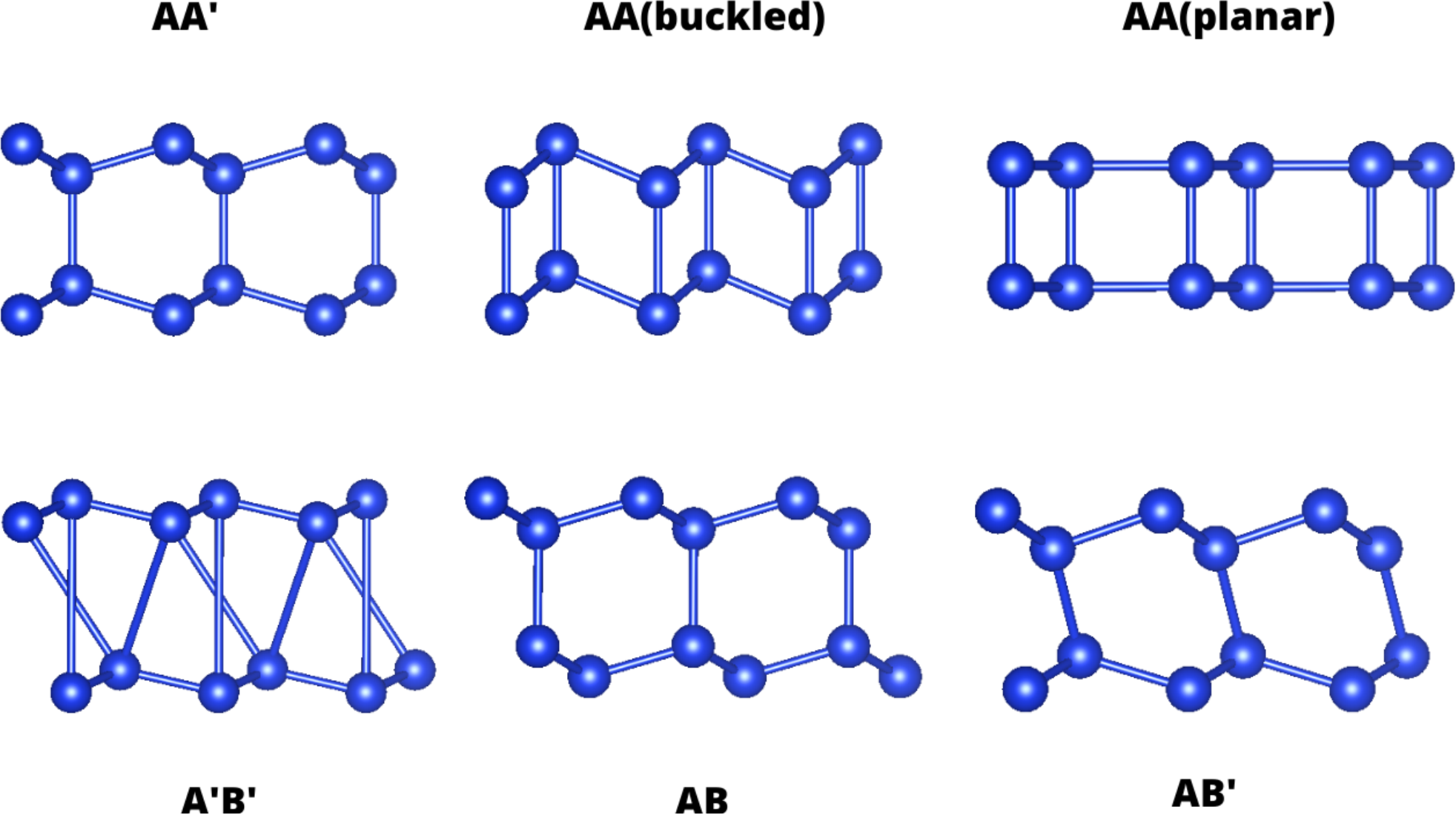}
     \caption{a) Top view and b) side view of various stackings in pristine bilayer silicene.}
     \label{fig:purebilayer}
 \end{figure*}

The relaxed structures are shown in Figure \ref{fig:purebilayer}. The structural properties of six stackings can be seen in Table \ref{tab:bilayer_geometry}. The AB, AB' and A'B' lattice constants and buckling are similar. However, the interlayer distance is very different. 

\begin{table}[!htb]
\centering
\begin{tabular*}{8cm}{@{\extracolsep{\fill}}lccc}
\hline
Stacking & $a$ (\AA) & $d$ (\AA) $\Delta$ (\AA) \\ \hline
AA'  & 3.86 & 2.45 & 0.66      \\ \hline
AAb  & 3.77 & 1.81 & 0.92      \\ \hline
AAp  & 4.12 & 2.40 & 0.0       \\ \hline
A'B' & 3.86 & 3.35 & 0.51      \\ \hline
AB   & 3.86 & 2.51 & 0.65      \\ \hline
AB'  & 3.85 & 2.19 & 0.68-0.77 \\ \hline
\end{tabular*}
\caption{Structural properties of pristine bilayer stackings: lattice parameter $a$, bond length $d$ and buckling  $\Delta$ (\AA).}
\label{tab:bilayer_geometry}
\end{table}

\begin{figure}[!htb]
    \centering
    \includegraphics[scale=0.45,width = 8cm]{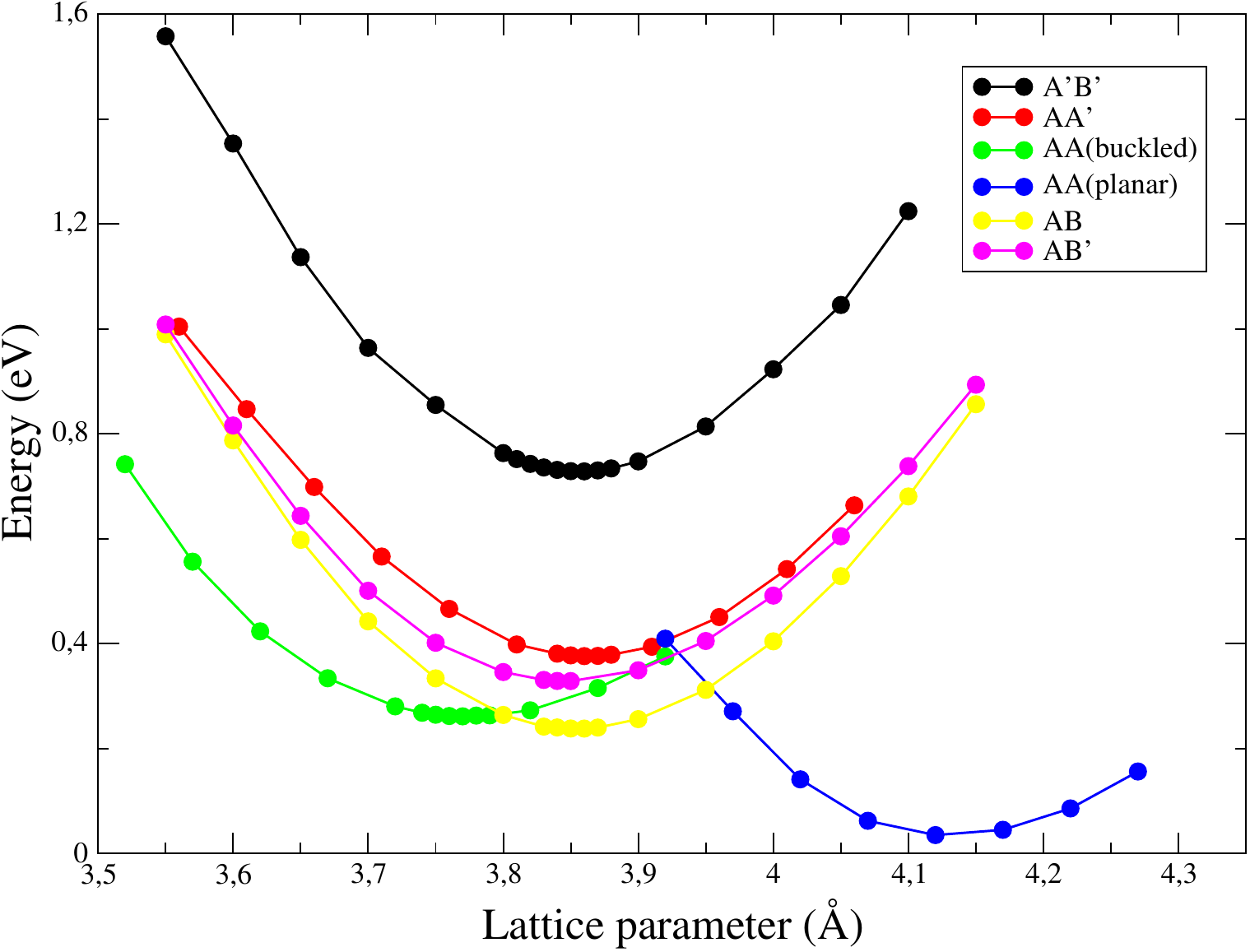}
    \caption{Energy as a function of the lattice parameter for bilayer stackings of pristine silicene.}
    \label{fig:purecomparision}
\end{figure}

Figure \ref{fig:purecomparision} shows the ground state energy difference among the bilayers as a function of the lattice parameter. The AA', AAb, AB and AB' have very similar energies. A'B' stacking mode is the less stable structure among all  free standing bilayers.  AAp  and AB are the most stable pristine bilayers  with the largest lattice constant, in agreement with Ref.\,\cite{artigo_33}. The band structure and density of states show that $p_z$ orbitals contribute the most in Fermi level $p_z$ orbitals are responsible to the bonding between the two layers, forming the bilayer. All the other stacking modes presents the same behavior.

\begin{figure}[!htb]
    \centering
   \includegraphics[scale=0.45, width = 6cm,clip]{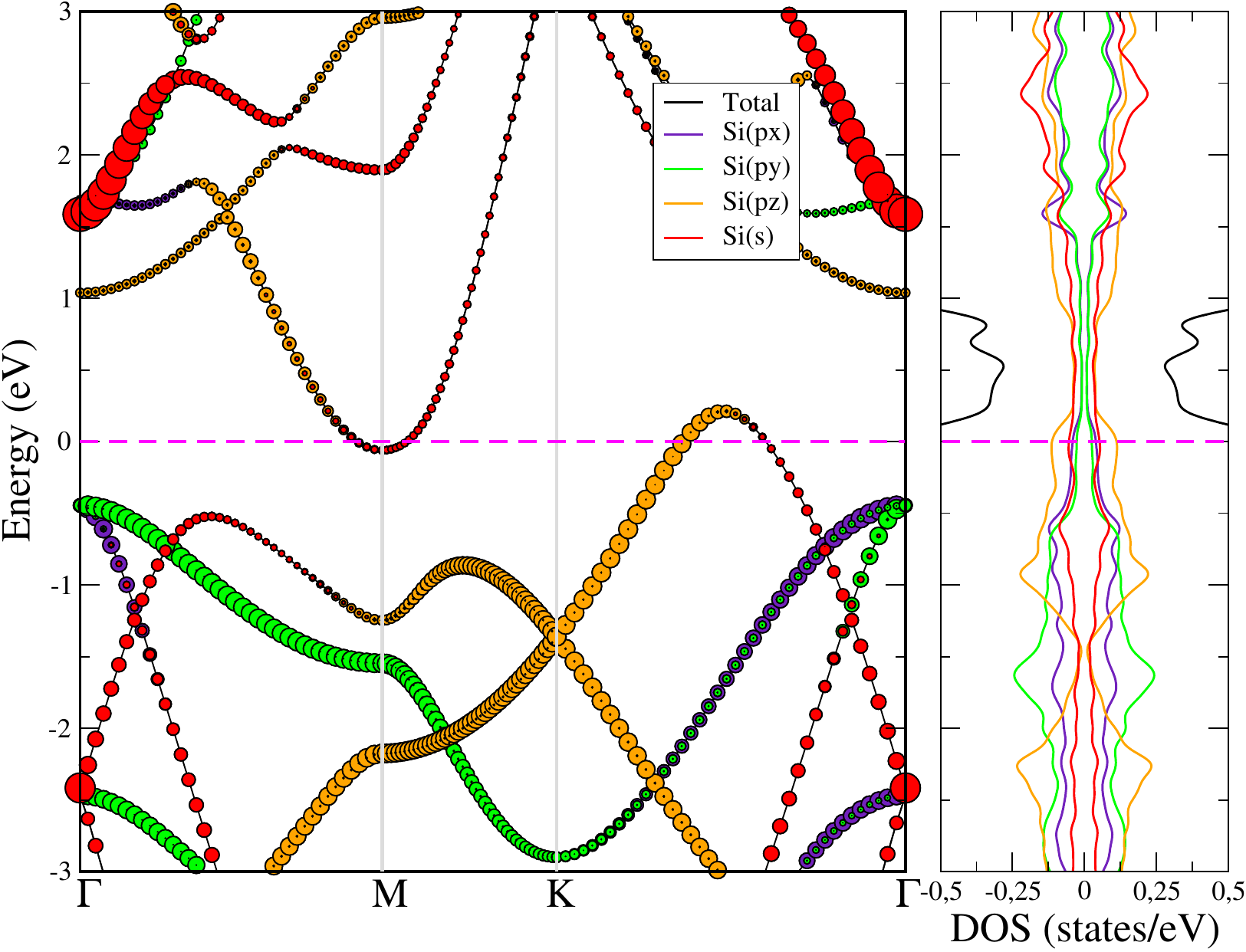}\\
   \includegraphics[scale=0.45, width = 6cm,clip]{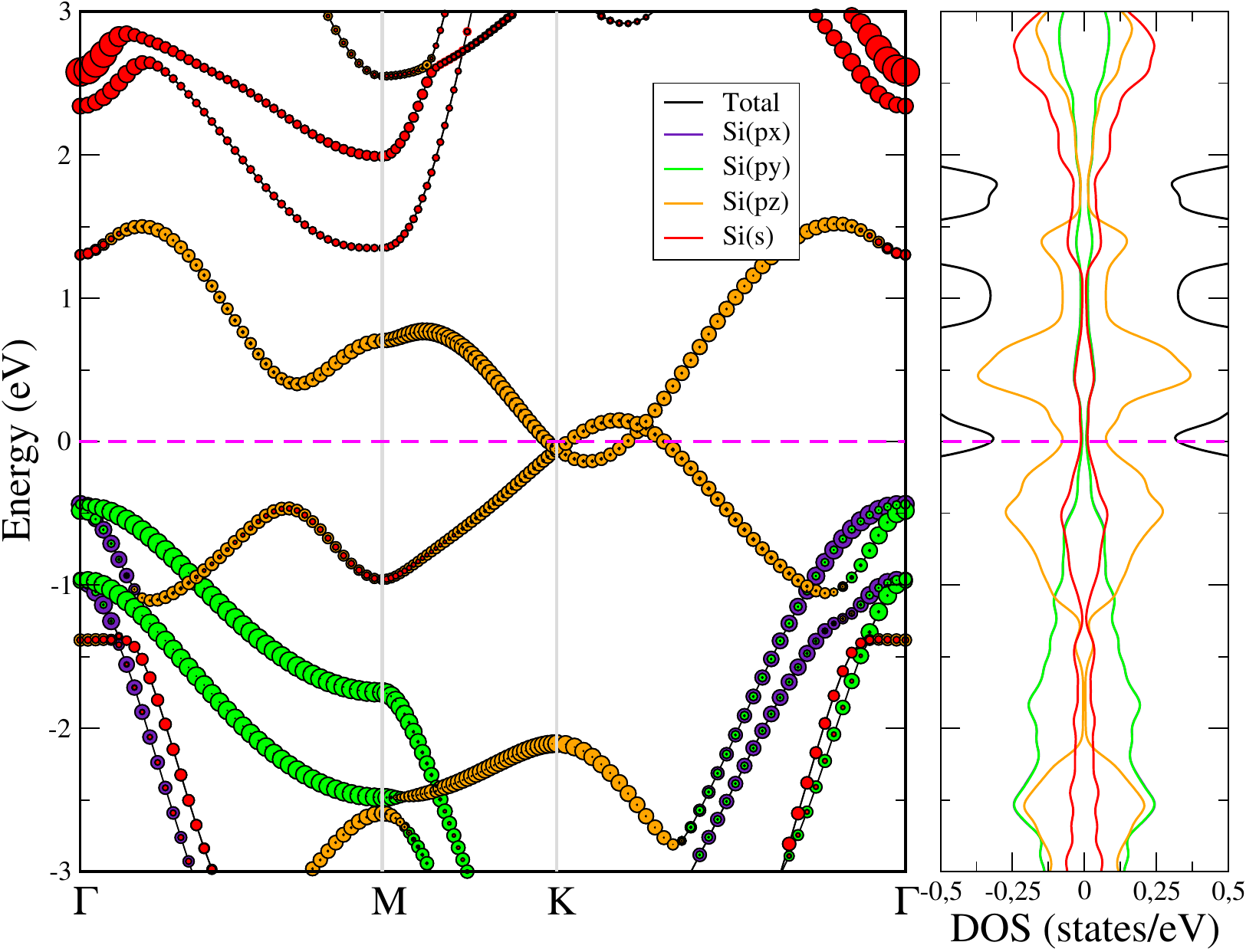}
    \caption{Band structure  and projected density of states of a) AA (buckled) and b) AB stacking in pristine bilayer silicene.}
    \label{fig:band_bilayer_bare}
\end{figure}

\section{Intercalation of Cr in silicene bilayers}
 
 The (1x1) unit cell of the stacking modes were doped with one Cr atom between the two layers of each kind of stacking shown in Figure \ref{fig:band_bilayer_bare}. It is obvious that all the structures had their properties modified by the presence of the metallic atom Table \ref{tab:properties_doped_bilayer}. All the structures had their interlayer  distance increased. The buckling of all structures changed. Structures such as  AA' and AB had their buckling increased, AAb was the only structure to have the buckling decreasing after being doped. The structure which had a increasing buckling tends to have a slightly smaller lattice constant. 

\begin{figure*}[!htb]
     \centering
 \includegraphics[scale=0.4, width = 12cm,clip]{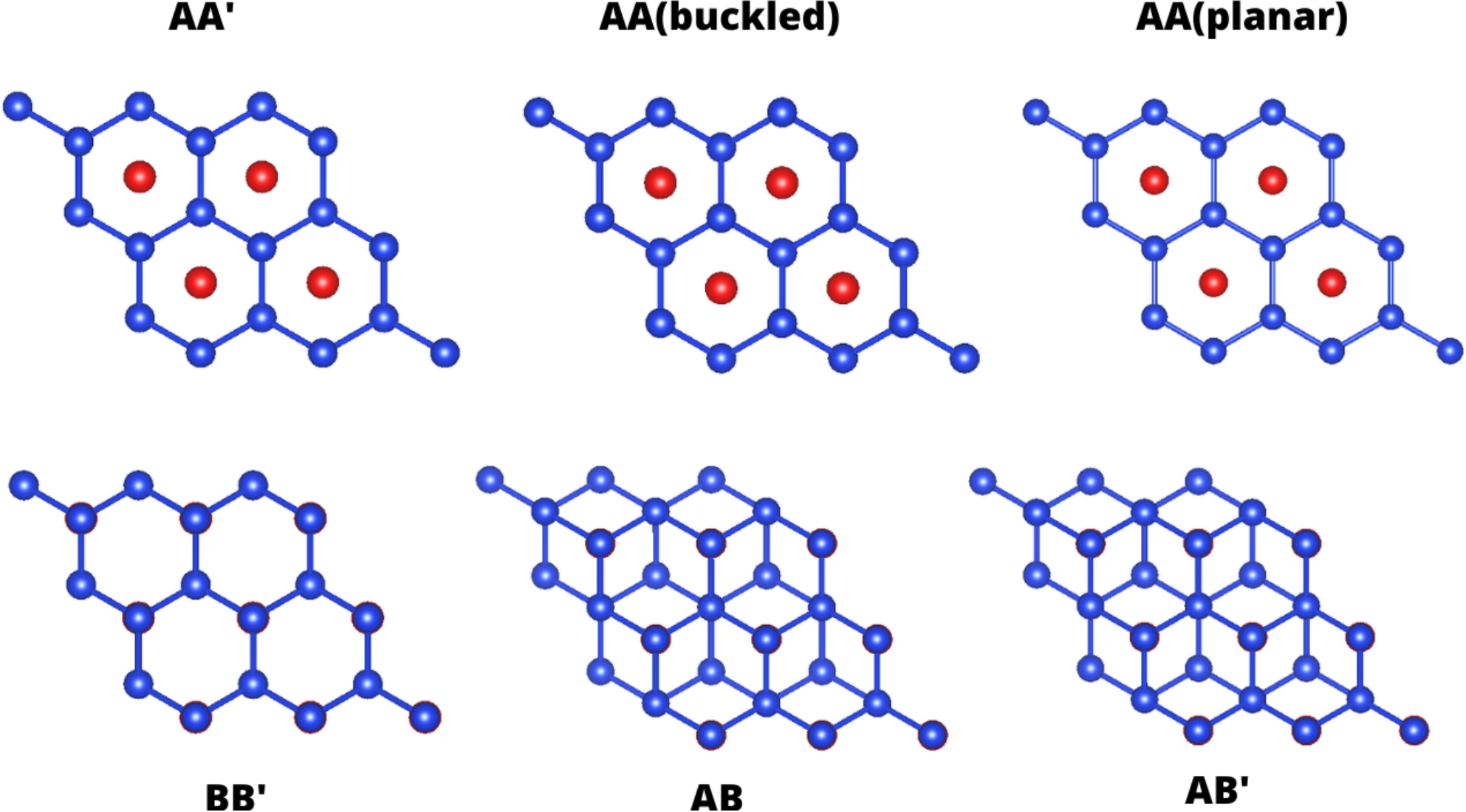}\\
 \includegraphics[scale=0.4,width = 12cm,clip]{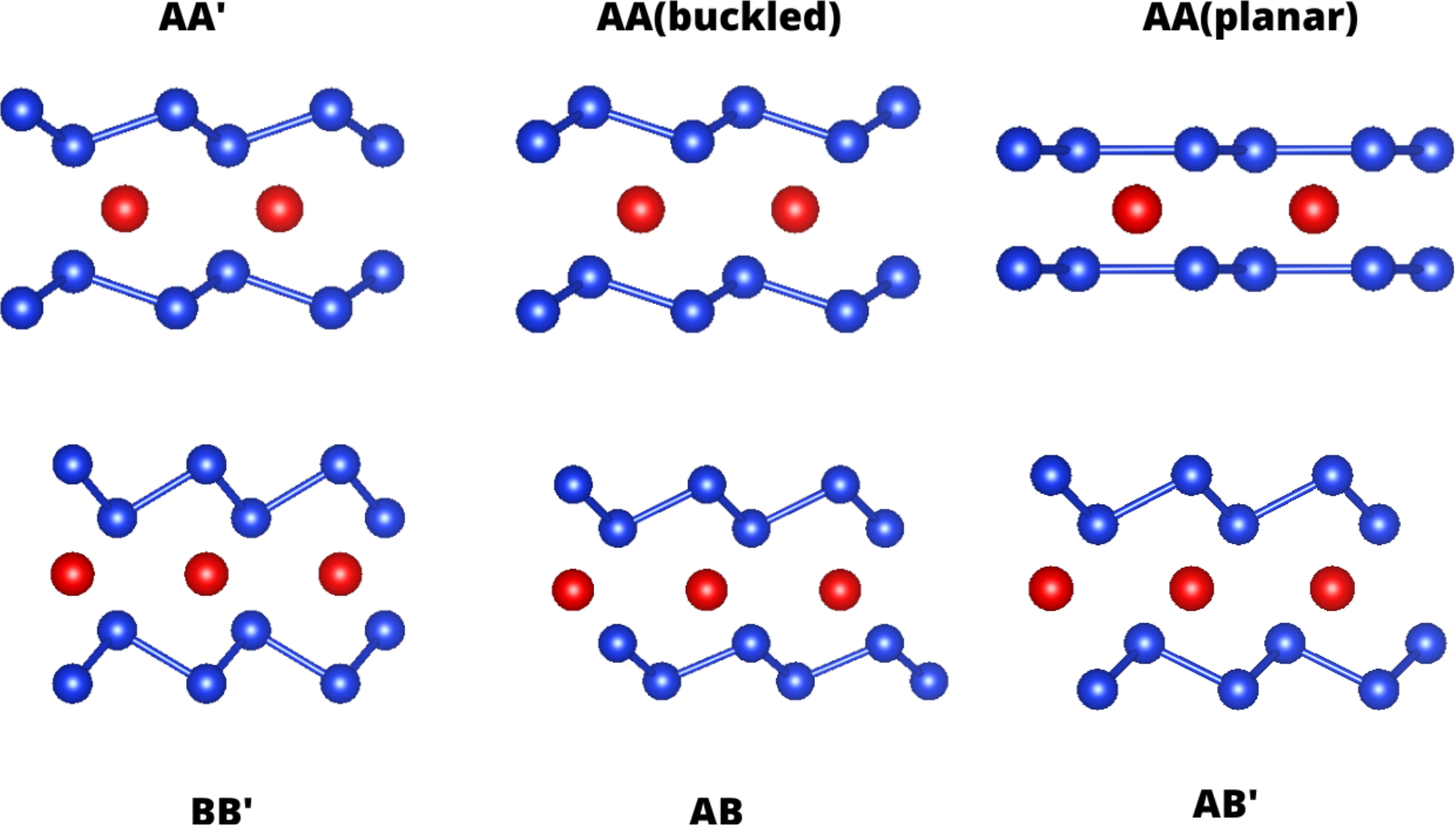}
     \caption{Top view and side view of  Cr doped bilayers in silicene.}
     \label{figure:doped_bilayer}
 \end{figure*}

  \begin{table}[!htb]
 \centering
 \begin{tabular*}{8cm}{@{\extracolsep{\fill}}lcccc}
\hline
Stacking & $a$ (\AA)&  $d$ (\AA) & $\Delta$ (\AA) & $\mu$ ($\mu_B$) \\ 
\hline
AA' & 3.86 & 2.72 & 0.78        & 3.68 \\ \hline
AAb & 3.82 & 2.90 & 0.73        & 3.43 \\ \hline
AAp & 4.22 & 2.46 & 0.0         & 3.84 \\ \hline
BB' & 3.61 & 2.62 & 1.25        & 1.25 \\ \hline
AB  & 3.76 & 2.80 & 0.92 - 1.07 & 3.18 \\ \hline
AB' & 3.70 & 2.71 & 1.05 - 1.12 & 2.35 \\ \hline
\end{tabular*}
\caption{Structural and magnetic properties of pristine bilayer stackings.}
\label{tab:properties_doped_bilayer}
\end{table}

\begin{figure}[!htb]
     \centering
     \includegraphics[scale=0.45,width = 7cm,clip]{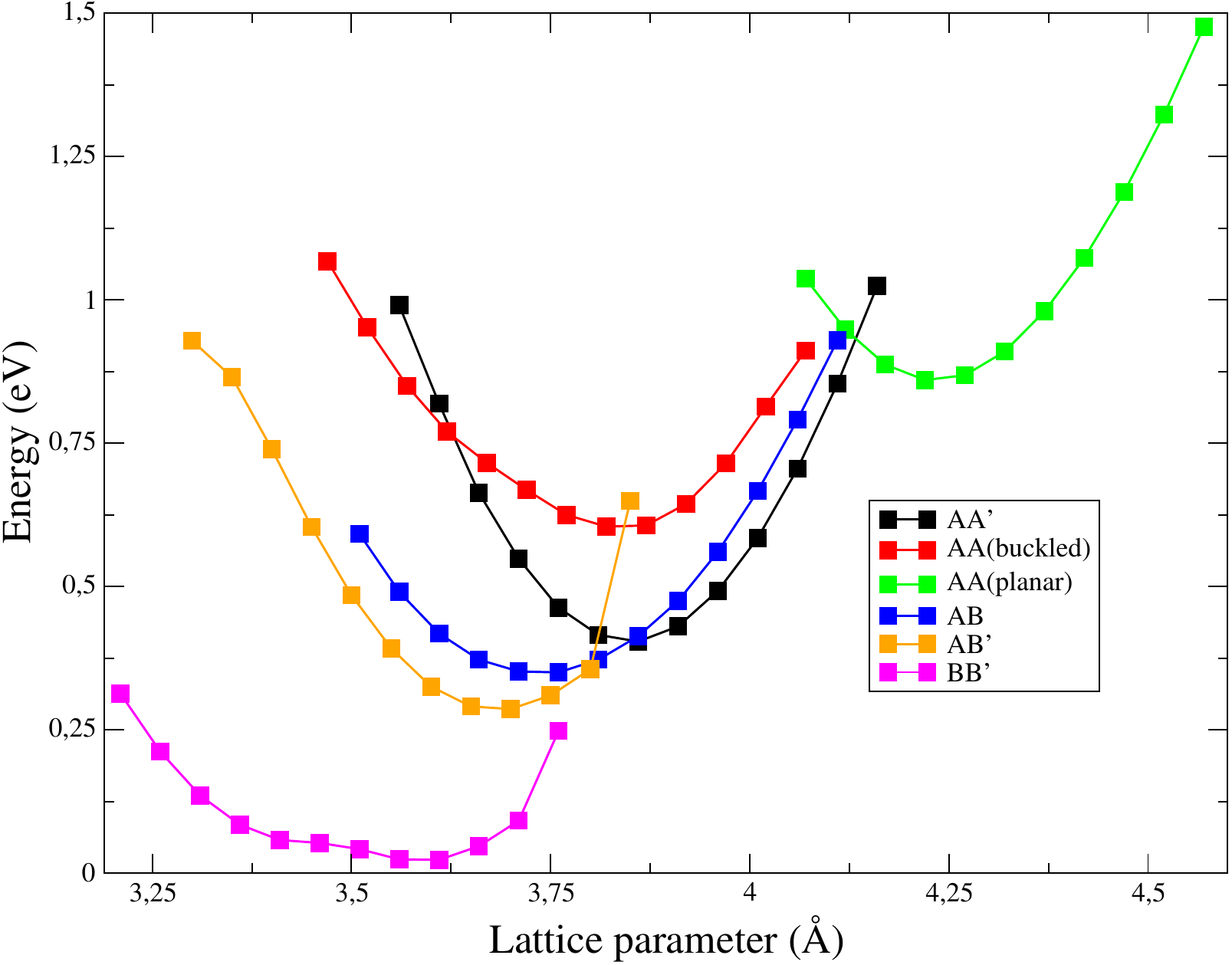}
     \caption{Energy as a function of the lattice parameter for stackings of Cr doped bilayer silicene.}
     \label{fig:doped_comparison_bilayer}
 \end{figure}

Upon doping AAp  becomes now the less stable structure. As a general behavior the doped hcp stacking modes are energetically more favourable compared to the remaining structures. One feature to be noticed is that the A'B' stacking mode have vanished and replaced by a staking mode called BB'. The ground state configuration for doped A'B' is probably metastable. The BB' stacking is the same as the AA' but the Cr atom was doped in the top site instead of the hollow site. Turns out that BB' was the most stable of all stacking modes, despite of having a flat energy ground state as can be seen in Figure \ref{fig:doped_comparison_bilayer}.

 The magnetic moment of doped bilayer silicene reaches the maximum in AAp (less stable) and reaches the minimum in BB' stacking mode (most stable). So it is possible to see  with the information of Table \ref{tab:properties_doped_bilayer} that there is relation between the magnetic moment and buckling of the bilayer. The larger the buckling the smaller is the magnetic moment. So it is important to analyse the density of states and the charge density distribution and associate this relation to the orbitals occupation at the Fermi level.

\begin{figure*}[!htb]
     \centering
  \includegraphics[scale=0.46,width=12cm,clip]{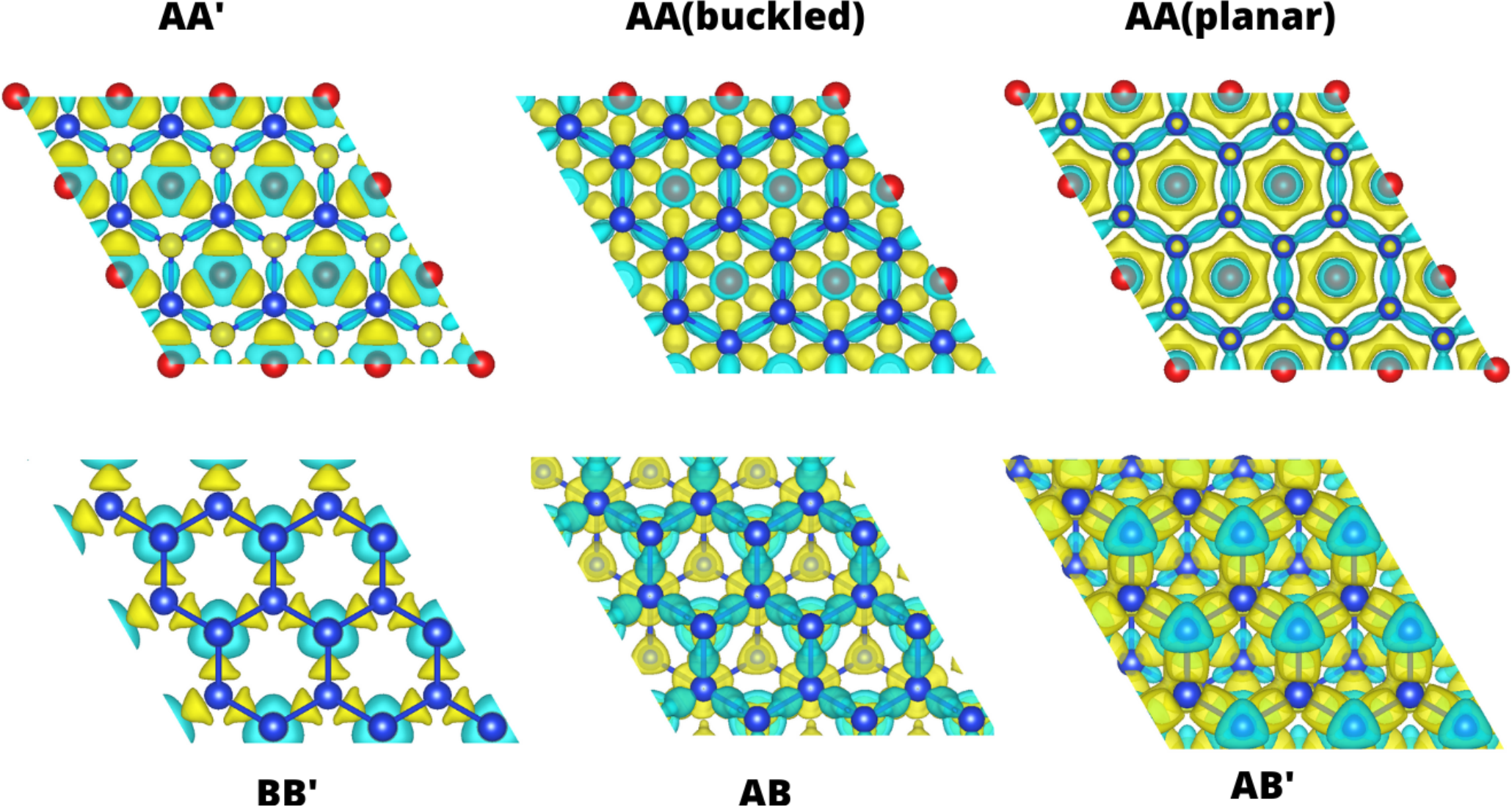}\\
   \includegraphics[scale=0.46,width=12cm,clip]{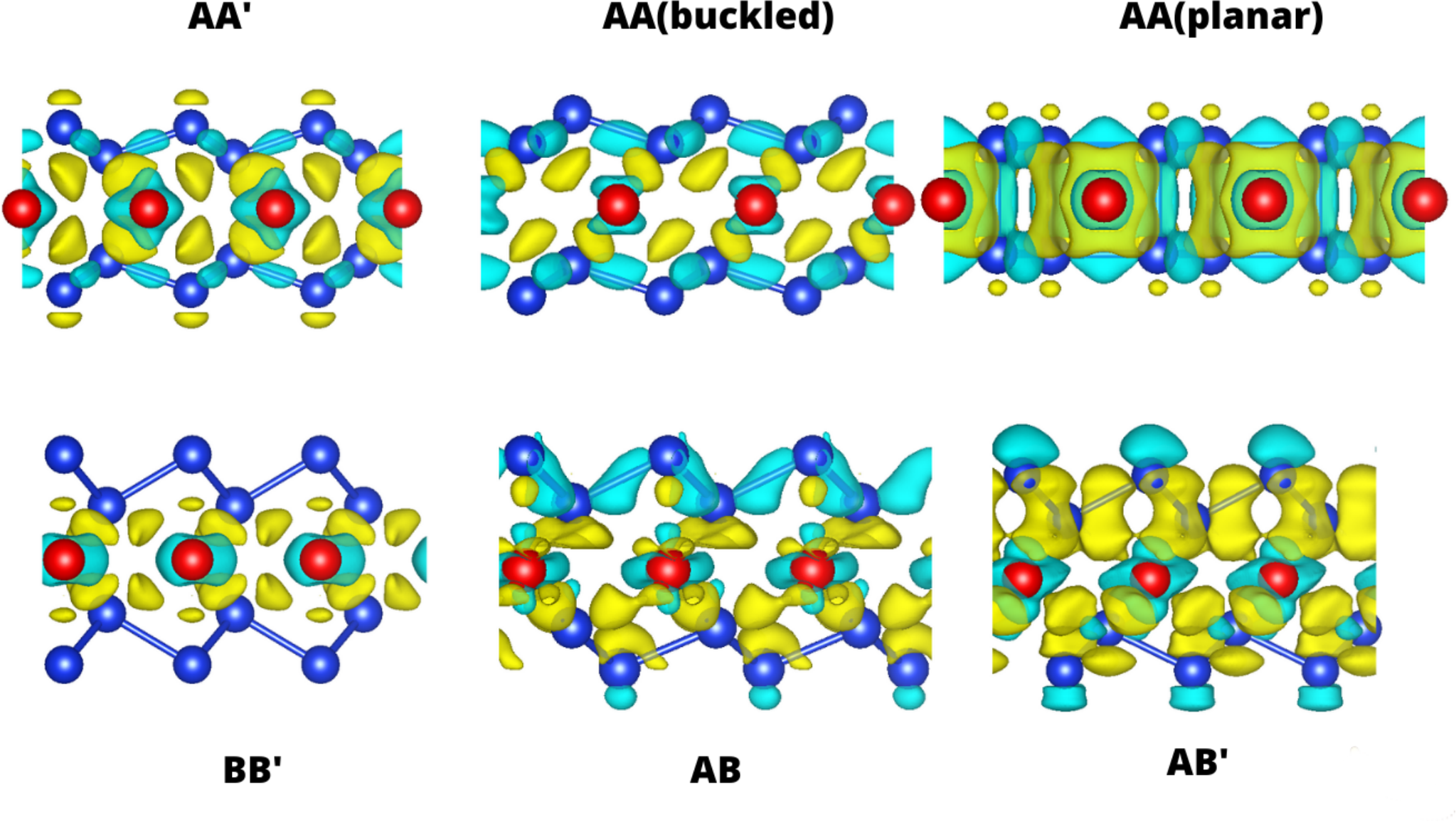}
     \caption{Charge density difference a) top view and b) side view of bilayer stacking modes. The yellow region is where the charge has accumulated and the cyan region is where there is a depletion of electrons.}
     \label{fig:charge_density_bilayer}
 \end{figure*}

 The charge density difference allows us to visualize  the charge density in the material after including a new atom in the structure.  One can obtain a image of the charge density difference as in Figures \ref{fig:charge_density_bilayer}.

 Figure \ref{fig:charge_density_bilayer} shows the charge density difference  where the yellow region is where the charge has accumulated and the cyan region is where there is a depletion of electrons. Therefore, one can easily see that in all  stacking modes the Si-Si bond in the same layer had a charge depletion, and the charge near the Cr atom have depleted too. It means that the silicene donated electrons to the Cr atom as seen in Figure \ref{fig:charge_density_bilayer}.  One can observe that the charge density have accumulated in a very specific shape between the Cr atom and the two layers. This shape indicates a formation of a new orbital, which comes from a hybridization of the $sp^3$ and $sp^2$  orbitals of silicene and the $d$ orbitals from Cr-$d$ orbital having a larger contribution at the Fermi level. It is possible to notice formation of six orbitals in every stacking mode. This happens because Cr atom electronic distribution in the valence level is $3d^54s^1$.

The charge density difference and the projected DOS show Si orbitals contribution at the Fermi level compared to the Cr orbitals contribution is very small, i.e., the potential of Si does not scatter strongly the Cr states, hence the orbitals are more localized as one can see in Figure \ref{fig:charge_density_bilayer}. Those features  make doped silicene bilayers good candidates for spintronics applications.

\begin{figure}[!htb]
\centering
    \includegraphics[scale=0.04,width = 6cm,clip]{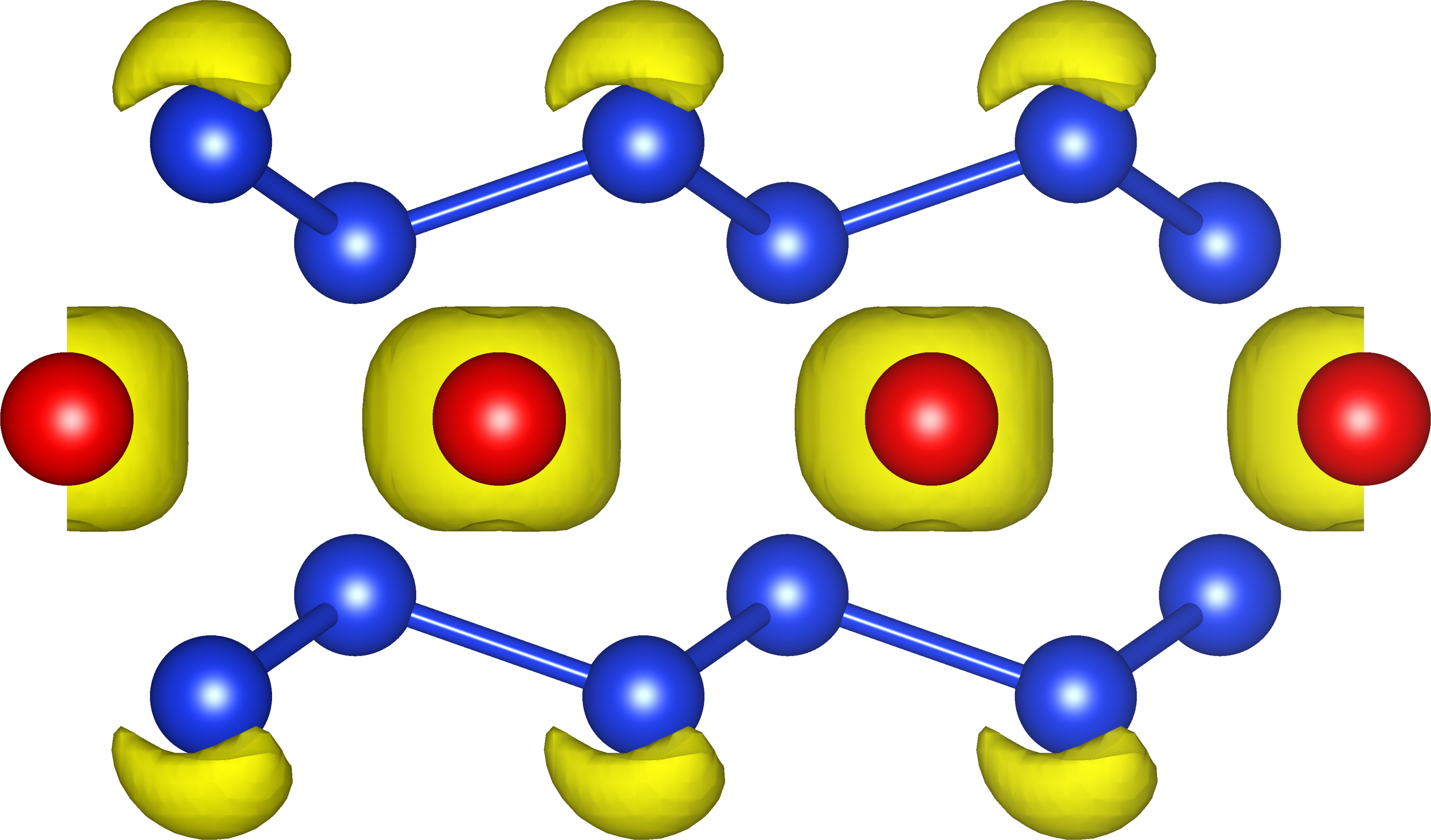}\\
    \includegraphics[scale=0.04, width = 6cm,clip]{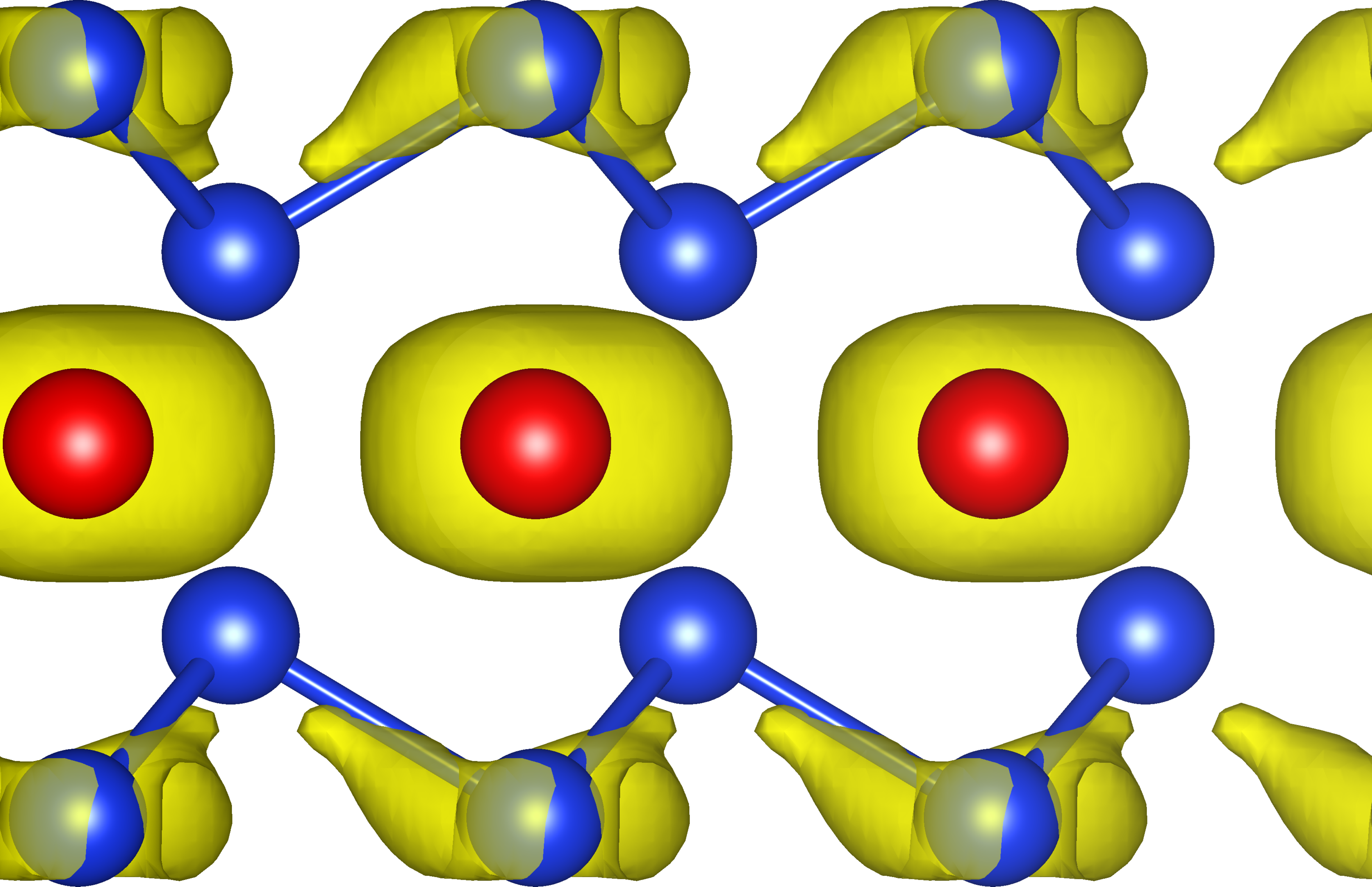}
    \caption{Partial charge at Fermi level for a) AA' and b) BB' stacking configurations.}
    \label{fig:partial_charge}
\end{figure}

\begin{figure}[!htb]
     \centering
     \includegraphics[scale=0.46,width = 7cm,clip]{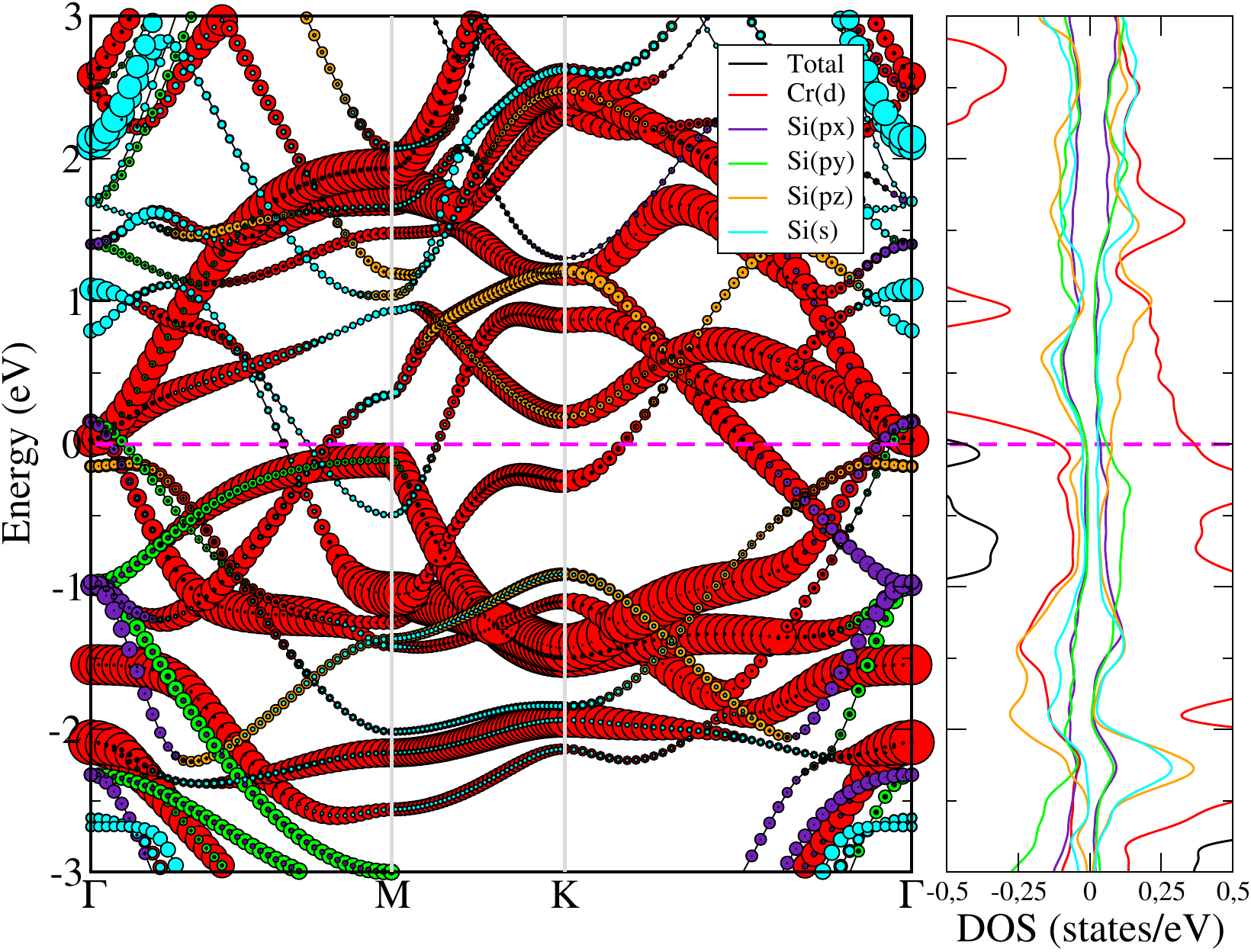}\\
      \includegraphics[scale=0.46, width = 7cm,clip]{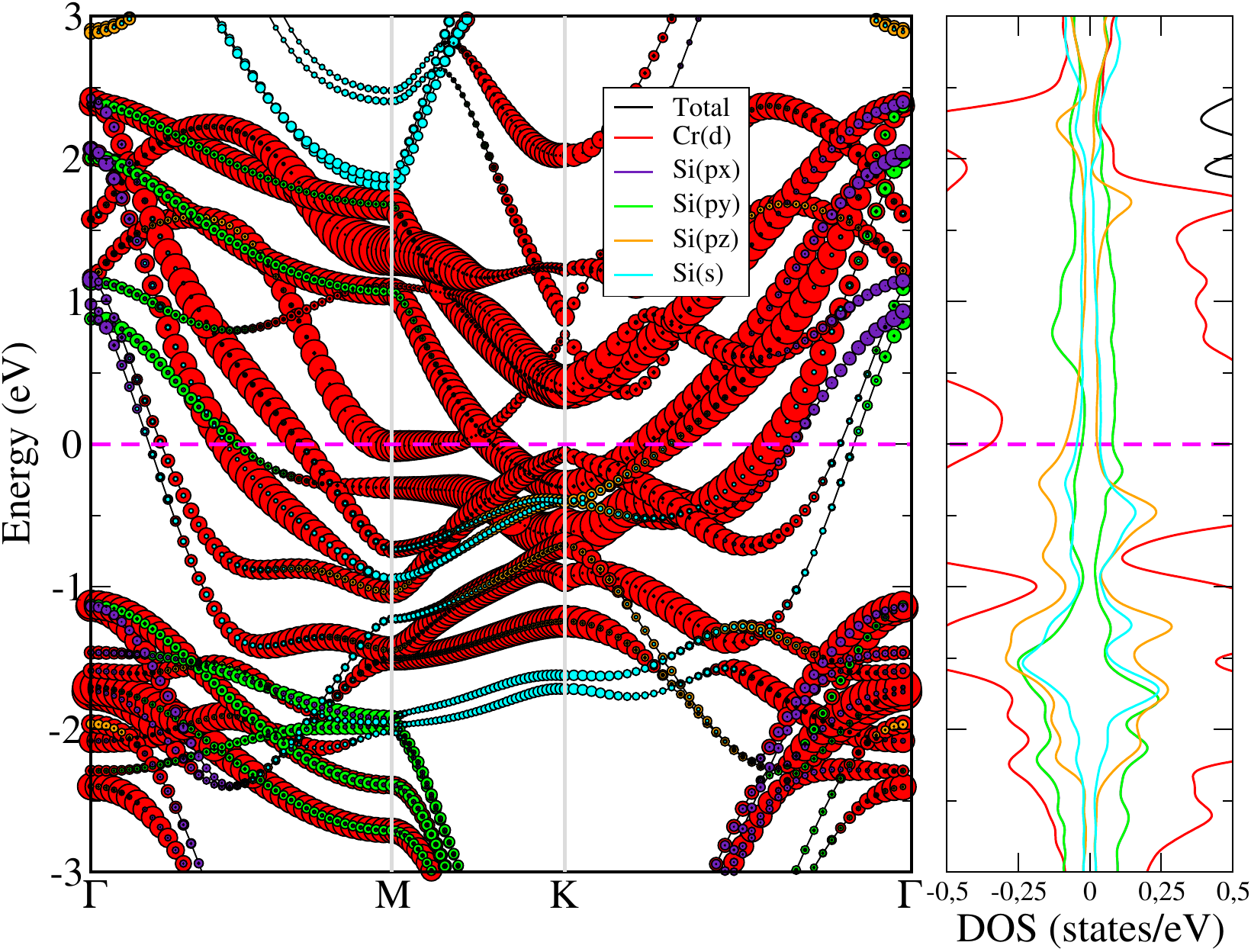}
     \caption{Projected band structure  and projected density of states of  Cr-doped a) AA' and b) BB' stacking mode.}
     \label{fig:band_bilayer_doped}
 \end{figure}

The larger is the buckling the smaller is the magnetic moment of the structure. In order to understand that, we take as showcases the  AA' and BB bilayers. Their charge density distribution close to the Fermi level  and the projected density of states are shown in Figures\,\ref{fig:partial_charge} and \ref{fig:band_bilayer_doped}, respectively.  We chose  AA' and BB', since AA' has a relatively low buckling a high magnetic moment. On the other hand we have BB' which has a high buckling and small magnetic moment. 

\begin{table*}[!htb]
\centering
\begin{tabular*}{12cm}{@{\extracolsep{\fill}}lccccccc}
\hline
\multicolumn{2}{c}{}    & \multicolumn{3}{c}{Low buckled}    \hspace{1cm}                                                 & \multicolumn{3}{c}{High Buckled}                               \\ \hline
\multicolumn{2}{c}{Stacking}       & \multicolumn{1}{c}{AA'}   & \multicolumn{1}{c}{AAb}   & \multicolumn{1}{c}{AAp}   \hspace{1cm}  & \multicolumn{1}{c}{AB}    & \multicolumn{1}{c}{AB'}   & BB'   \\ \hline
\multicolumn{2}{c}{\begin{tabular}[c]{@{}c@{}}\\ $\mu(\mu_B)$\end{tabular}}                                                  & \multicolumn{1}{c}{3.68}  & \multicolumn{1}{c}{3.43}  & \multicolumn{1}{c}{3.84}   \hspace{1cm} & \multicolumn{1}{c}{1.27}  & \multicolumn{1}{c}{3.18}  & 2.35  \\ \hline
\multicolumn{1}{c}{\multirow{2}{*}{\begin{tabular}[c]{@{}c@{}}spin up \\ \end{tabular}}}   & \multicolumn{1}{c}{$p_x$} & \multicolumn{1}{c}{0.036} & \multicolumn{1}{c}{0.054} & \multicolumn{1}{c}{0.019}  \hspace{1cm} & \multicolumn{1}{c}{0.098} & \multicolumn{1}{c}{0.094} & 0.080 \\ \cline{2-8} 
\multicolumn{1}{c}{}                                                                                              & \multicolumn{1}{c}{$p_y$} & \multicolumn{1}{c}{0.059}   & \multicolumn{1}{c}{0.054} & \multicolumn{1}{c}{0.042}  \hspace{1cm} & \multicolumn{1}{c}{0.098} & \multicolumn{1}{c}{0.094} & 0.080 \\ \hline
\multicolumn{1}{c}{\multirow{2}{*}{\begin{tabular}[c]{@{}c@{}}spin down \\ \end{tabular}}} & \multicolumn{1}{c}{$p_x$} & \multicolumn{1}{c}{0.014} & \multicolumn{1}{c}{0.061} & \multicolumn{1}{c}{0.059}  \hspace{1cm} & \multicolumn{1}{c}{0.028} & \multicolumn{1}{c}{0.014} & 0.028 \\ \cline{2-8} 
\multicolumn{1}{c}{}                                                                                              & \multicolumn{1}{c}{$p_y$} & \multicolumn{1}{c}{0.014} & \multicolumn{1}{c}{0.061} & \multicolumn{1}{c}{0.067}  \hspace{1cm} & \multicolumn{1}{c}{0.028} & \multicolumn{1}{c}{0.014} & 0.028 \\ \hline
\end{tabular*}
\caption{Silicon $p_x$ and $p_y$ orbitals contribution in states/eV at the Fermi level of stacking modes.}
\label{tab:orbital_bilayer}
\end{table*}

Figure \ref{fig:partial_charge} shows  that the charge distribution near the Fermi level of the two stacking modes are different.Figure \ref{fig:partial_charge} shows an accumulated charge in the silicene monolayers and all around the Cr atom. So in AA' stacking mode shown in  Figure \ref{fig:partial_charge} it is possible to see the charge in the outer layers of the silicene monolayer is related to the $p_z$ orbital. The BB' stacking mode shown in Figure \ref{fig:partial_charge} is easy to identify that the charge is accumulating between the Si-Si bonds, the accumulated charge are $p_x$ and $p_y$ orbital.

In Table \ref{tab:orbital_bilayer} we show the silicon orbitals contribution to the hybridization with Cr-$d$ orbitals near the Fermi level. It is possible to notice that Si $p_x$ and $p_y$ orbitals have smaller contributions in low buckled stacking modes to the hybridization with Cr-$d$ orbitals. On the other hand the contribution of Si $p_x$ and $p_y$ orbitals have larger contributions in  the high buckled stacking modes. Therefore, one can relate the increase of the magnetic moment of to the density contribution of Si $p_x$ and $p_y$ in the hybridization with Cr-$d$ orbitals. This relation can be seen by comparing the AA' and BB' stacking modes. A similar behavior is found for Cr-doped dumbbell silicene. Moreover, one can see that the charge accumulated around the Cr atom is more localized for the AA' compared to the BB' stacking. Notice that the layer distance in the two stacking modes are almost the same.

\section{Conclusions} 

 The adsorption of Cr atom on silicene  changes significantly the structural properties and electronic properties of the monolayer. The adsorption or substitution of Cr atom often causes a strain on silicene yielding a increase not just in the buckling but also in the lattice constant. Moreover, magnetic properties arise in all doped Si based structures. As the concentration of Cr atoms adsorbed or substituted increase, the magnetic moment of the system tends to decrease. Diffusion barrier calculations show that if an Cr atom is adsorbed in silicene, it is very unlikely that the Cr atom moves for an adjacent hollow site, or cross the silicene structure since the diffusion barriers are very high. The magnetic properties study of Cr pair adsorbed on silicene showed that arm-chair direction is usually more stable configuration than zig-zag direction of adsorption. Low concentrations of adsorbed Cr atom pair in arm-chair direction tends to a AFM state. But in the majority of the systems studied their is no big energy difference between  the FM and AFM states, so strain controlled could be applied for spintronic applications. Dumbbell silicene adsorbed with Cr atoms pair is more stable in AFM configuration. The adsorption of the Cr pair changes strongly the pristine dumbbell silicene structural and electronic properties. The Cr doped bilayer showed a very interesting relation between the buckling of silicene and the total magnetic moment of the system. The larger is the buckling the smaller is the magnetic moment. Using the charge density analysis and projected DOS one could relate the variation in the magnetic moment to the contribution of Si $p_x$ and $p_y$ orbitals to the hybridization with Cr $d$ orbitals. Cr doped bilayer silicene could be a promising material for spintronic applications since good qubits is found in the stacking modes studied.

\section{Acknowledgements}

We acknowledge the financial support from the Brazilian funding agency CNPq under grant numbers 313081/2017-4 and 305335/2020-0. L. M. G. thanks for a CNPq fellowship. We thank computational resources from LaMCAD/UFG, Santos Dumont/LNCC, CENAPAD-SP/Unicamp.

 \bibliographystyle{apsrev4-1} 
!
\end{document}